\documentclass[12pt,fleqn]{article}
\pdfoutput=1
\usepackage{amsmath}
\usepackage{amssymb} % e.g. \gtrsim
\usepackage{bbm} % BlackBoard letters
\usepackage[small]{caption2} % font of captions is different from text
\usepackage{graphicx} % \including PostScript
\usepackage{mathrsfs}   % e.g. nice Lagrange-L with \mathscr{L}
\usepackage{cite} % [1-5] instead of [1,2,3,4,5]
\usepackage{color}

\usepackage{slashed}
\usepackage{url}

\usepackage[pdftex,hyperfootnotes=false,pdfpagelabels]{hyperref}

\newcommand{\eVdist}{\kern-0.06em}
\newcommand{\gev}{\:\text{Ge\eVdist V}}

\newcommand{\beq}{\begin{eqnarray}}
\newcommand{\eeq}{\end{eqnarray}}

\allowdisplaybreaks[1]

\begin{document}

\begin{titlepage}

\vspace*{-3.0cm}
\begin{flushright}
OUTP-12-05P
\end{flushright}

\begin{center}
{\large\bf LHC and Tevatron bounds on the dark matter direct detection
  cross-section for vector mediators }

\vspace{1cm}

\textbf{
Mads T.~Frandsen,
Felix Kahlhoefer,
Anthony Preston,\\
Subir Sarkar,
Kai Schmidt-Hoberg
}
\\[5mm]
\textit{\small
Rudolf Peierls Centre for Theoretical Physics, University of Oxford,\\
1 Keble Road, Oxford OX1 3NP, UK
}
\end{center}

\vspace{1cm}

 \begin{abstract}
 We study the interactions of a new spin-1 mediator that connects the
 Standard Model to dark matter. We constrain its decay channels
 using monojet and monophoton searches, as well as searches for
 resonances in dijet, dilepton and diboson final states including 
 those involving a possible Higgs. We then interpret the
 resulting limits as bounds on the cross-section for dark matter
 direct detection without the need to specify a particular model.  For
 mediator masses between 300 and 1000~GeV these bounds are
 considerably stronger than the ones obtained under the assumption
 that the mediator can be integrated out. 
\end{abstract}

\end{titlepage}

\tableofcontents
\newpage

%%%%%%%%%%%%%%%%%%%%%%%%%%%%%%%%%%%%%%%%%%%%%%%%%%%%%%%%%%%%%%%%%%%%%%%%

\section{Introduction}

Although the Large Hadron Collider (LHC) has yet to see any evidence
for physics beyond the Standard Model (SM), we know from astrophysical
and cosmological observations that the SM is incomplete, because it
lacks an adequate candidate for the dark matter (DM)
particle. Moreover, there are likely to be new interactions that
connect the DM particle to the SM. In this paper, we discuss how
colliders~-- and especially the LHC~-- can contribute to the search
for DM and such new interactions.

The search for dark matter has mainly been focused on particles with
mass and interactions set by the Fermi scale, e.g.\ neutralinos in
extensions of the SM with softly broken supersymmetry (SUSY) at this
scale. With an impressive increase of sensitivity over the past few
years, direct detection experiments such as XENON100
~\cite{Aprile:2011hi} and CDMS-II~\cite{Ahmed:2009zw,Ahmed:2010wy} have
pushed down upper bounds on the (spin-independent) scattering
cross-section on nuclei for such particles and have begun to
constrain the relevant SUSY parameter space. 

For much lighter
DM particles, however, direct detection bounds become significantly weaker and
at the same time such particles must have stronger interactions if
their relic annihilations are to result in an acceptable DM
abundance. Hence an improvement of the bounds for light DM is of great
interest especially since recent results from the
DAMA~\cite{Bernabei:2010mq}, CoGeNT~\cite{Aalseth:2011wp} and
CRESST-II~\cite{Angloher:2011uu} experiments hint at DM with mass $m_\chi
\sim 10$~GeV and cross-section $\sigma_p \sim 10^{-40} \text{
  cm}^2$. If DM does have such properties, then production of DM pairs
at the LHC would be sizeable and result in a variety of
observable signals.

Among the most promising signatures of DM at high energy colliders are
excesses of events with either a single high-energy jet or a single
high-energy photon and a large amount of missing transverse
energy (MET). Such monojet and monophoton searches have been
performed at LEP, Tevatron and the LHC but no excess has been observed
over expected SM backgrounds~\cite{Abdallah:2003np, CDF-monojet,
  CMS-monojet, Aad:2011wc, CMS-Monophoton1, CMS-Monophoton2}. If the mediator of the
DM interaction with the SM is so heavy that it cannot be
produced on-shell at the LHC, then these searches directly bound the coupling
of DM to nucleons and are competitive with the bounds set by direct
detection experiments~\cite{Goodman:2010yf,Bai:2010hh,
  Rajaraman:2011wf,Goodman:2010ku,Fox:2011fx, Fox:2011pm, Shoemaker:2011vi,
  Fox:2012ee,MarchRussell:2012hi}. However, for such large mediator masses, the relevant
cross-sections for both the LHC and direct detection experiments will
be very small unless the coupling constants approach the bounds from
perturbative unitarity~\cite{Shoemaker:2011vi,Fox:2011pm,Fox:2012ee}.
  
Of course the mediator mass may well be comparable to LHC energies. In
this case an effective operator description is no longer valid,
because the LHC can resolve the interaction and produce the mediator
on-shell, which complicates the comparison with direct
detection experiments. On the plus side, it opens up the possibility
to search for resonances in various channels from the
decays of the new mediator into SM particles. Combining the limits from
all relevant collider searches, it is still possible to constrain the direct
detection cross-section in a \emph{model independent} way.

In particular, if DM is light compared to the dominant mediator, we can bound the DM direct detection cross-section in terms of the total width and the invisible branching ratio of the mediator in a simple way. These two quantities may in turn be constrained by collider searches without having to specify an underlying model~-- even though the collider bounds may be much stronger in a specific model framework.
In this paper we apply this approach, assuming that the interaction between DM and the SM is
dominated by the exchange of a neutral spin-1 state, here termed
$R$.  An example of such a spin-1 state is the $Z'$
associated with a new broken $U(1)$
symmetry~\cite{Holdom:1985ag,Babu:1997st,Cassel:2009pu,Hook:2010tw,Mambrini:2010dq,Kang:2010mh,Chun:2010ve,Fox:2011qd,Mambrini:2011dw,Gondolo:2011eq,Mambrini:2011pw,Frandsen:2011cg,Cline:2011zr,Heeck:2011md}. Another
example is a new resonance associated with a strongly interacting
extension of the SM~\cite{Foadi:2008qv,Barbieri:2010mn}, e.g.\ an
analogue of the neutral isospin zero $\omega$ resonance in QCD. 
Recently, there
have been analyses of the `dark Higgs' associated with the
$Z'$~\cite{Weihs:2011wp} and LHC signatures of a baryonic
$Z'$~\cite{An:2012va}.

The outline of this paper is as follows: In Section~\ref{sec:theory} we
introduce an effective Lagrangian for the spin-1 state, discuss the
decay channels and present the production cross-sections at LHC and
the Tevatron.  In Section~\ref{sec:bounds} we first summarize the
resulting collider bounds from LHC and Tevatron on the various decay modes of $R$, before discussing the bound for each decay mode in more detail. In
Section~\ref{sec:directdetection}, we compare our results to limits from direct detection on spin-independent and spin-dependent interactions. We do this both model independently and in the framework of a spin-1 state coupling to the SM via kinetic and mass mixing only, as discussed above.
A discussion of all assumptions
and their validity is given in Section~\ref{sec:discussion} together
with our conclusions. Appendices~\ref{sec:widths}
and~\ref{sec:couplings} provide, respectively, all relevant formulae
for the partial decay widths and the coupling constants of $R$.

\section{Interactions of a neutral spin-1 mediator}
\label{sec:theory}

We start from an effective Lagrangian (similar to ~\cite{Chun:2010ve})
describing the interactions of the neutral spin-1 state $R$ with the
SM fields and the DM particle. We then discuss all possible decay
channels and describe how they can be constrained by collider
searches.

\subsection{Effective Lagrangian description}

We divide the Lagrangian into the couplings to DM, SM fermions, SM
gauge bosons, the Higgs, and anything else
\beq
\mathcal{L}^{R}=\mathcal{L}^{R}_{\rm DM}+\mathcal{L}^{R}_{
  f\bar{f}}+\mathcal{L}^{R}_{\rm gauge}+\mathcal{L}^{R}_{H} +
\mathcal{L}^{R}_X \ .  
\eeq 
Depending on whether DM is a Dirac fermion
$\chi$ or complex (pseudo-)scalar $\phi$, we define $\mathcal{L}^{R}_{\rm
  DM}\equiv \mathcal{L}^{R}_\chi$ or $ \mathcal{L}^{R}_{\rm DM} \equiv
\mathcal{L}^{R}_\phi$ as appropriate, where 
\beq
\mathcal{L}^{R}_\chi=R_\mu \bar{\chi} \gamma^\mu (g^V_{\chi
  R}-g^A_{\chi R}\gamma^5) \chi \ , \quad \mathcal{L}^{R}_\phi=
g_{\phi R} R_\mu J^\mu_\phi 
\eeq 
and $J^\mu_\phi\equiv i (\phi^*
\partial^\mu \phi - \phi \partial^\mu \phi^*)$. We will not consider
the $CP$-odd operator $R^\mu \partial_\mu (\phi^*\phi)$ here 
(for a discussion see e.g.~\cite{DelNobile:2011uf}).

The interactions of $R$ with the SM fermions are described by
\beq
\mathcal{L}^{R}_{ f\bar{f}}= \sum_{f=q, \ell, \nu }R_\mu \bar{f} \gamma^\mu (g^V_{f R}-g^A_{f R}\, \gamma^5)f  \;,
\eeq
where $q,\ell,\nu$ denote SM quarks, charged leptons and neutrinos respectively.

Neglecting $CP$-violating terms (see e.g.~\cite{Hagiwara:1986vm} for a
more complete discussion) the couplings of $R$ to SM gauge fields can
be written as
\beq
\mathcal{L}^{R}_{\rm gauge} &=& g^{R}_{WW1} [[RW^+W^-]]_1 + g^{R}_{WW2} [[RW^+W^-]]_2 \nonumber \\ 
& + & g^{R}_{ZWW1} ((R Z W^+W^-)) + g^{R}_{\gamma WW1} ((R \gamma W^+W^-))  \nonumber \\
& + & g^{R}_{ZZ} [[R ZZ]]_\epsilon+ g^{R}_{Z\gamma} [[RZ\gamma]]_\epsilon + g^{R}_{WW3} [[RW^+W^-]]_\epsilon \nonumber \\
& + & g^{R}_{ZWW2}  \epsilon^{\mu\nu\rho\sigma} R_\mu Z_\nu W_\rho^{+} W_\sigma^{-} + g^{R}_{\gamma WW2} \epsilon^{\mu\nu\rho\sigma} R_\mu \gamma_\nu W_\rho^{+} W_\sigma^{-} \; ,
\label{eq:gauge}
\eeq
where
\beq
[[RW^+W^-]]_1 & \equiv & i\left[(\partial_\mu W_\nu^+ - \partial_\nu W_\mu^+) W^{\mu -} R^{\nu} - (\partial_\mu W_\nu^- - \partial_\nu W_\mu^-) W^{\mu +} R^{\nu}\right] \;, \nonumber \\{}
[[RW^+W^-]]_2 & \equiv & \frac{i}{2}(\partial_\mu R_\nu - \partial_\nu R_\mu) (W^{\mu +} W^{\nu -} - W^{\mu -} W^{\nu +})  \;,\nonumber\\{}
[[RV_1V_2]]_\epsilon & \equiv & \epsilon^{\mu\nu\rho\sigma}   (V_{1 \mu} \partial_\rho V_{2 \nu}- \partial_\rho V_{1\mu}  V_{2\nu})  R_\sigma \;, \nonumber \\
((RVW^+W^-)) & \equiv &  2 R_\mu V^\mu  W_\nu^{-} W^{\nu +} - R_\mu W^{\mu +}  V_\nu W^{\nu -} - R_\mu W^{\mu -}  V_\nu W^{\nu +}\nonumber \; ,
\eeq
for appropriate combinations of $V_{i}=\{\gamma,Z, W^+, W^- \}$.  
The operators in the first two lines of Equation~(\ref{eq:gauge})
conserve $C$ and $P$ separately, while the operators in the last two
lines are $CP$ even but $P$ odd. If the underlying theory violates $CP$,
then $CP$ violating couplings are also possible~\cite{Keung:2008ve}. In
principle there could also be a coupling between the $R$ and two
photons, leading to the decay $R\rightarrow \gamma^*
\gamma$.\footnote{Note that the Landau-Yang theorem only applies to
  on-shell photons.}  Such a decay would however correspond to a
non-renormalisable operator which~-- in order to have any observable
effects~-- would imply additional new physics at rather low scales. We
will not discuss such operators.

For comparable couplings 
the triboson final states are suppressed compared to the diboson
ones due to smaller available phase-space 
This is expected when $R$ is a new gauge boson (for details see e.g.~\cite{Rizzo:1989pi}).
Consequently, we will neglect triboson
decays of $R$ in the following.\footnote{However if $R$ is a
  `techni-omega', the coupling to the triboson final states can be
  enhanced so that the corresponding decays contribute significantly
  to the total width of $R$.}

Finally, couplings of $R$ to the SM Higgs are of the form
\beq
\mathcal{L}^{R}_{H}&=& g^R_{ZH} R_\mu Z^\mu H + g^R_{ZHH} R_\mu Z^\mu H^2 \; .
\eeq
Again, we expect decays into $ZHH$ to be significantly suppressed
compared to decays into $ZH$ so we neglect them in the following. A
coupling of $R$ to $HH$ is absent because $H$ is a real scalar and
terms proportional to $\partial_\mu R^\mu$ are $CP$ violating hence
we neglect them~\cite{Hagiwara:1986vm}.

Leaving $\mathcal{L}^{R}_X$ unspecified for now, the decay modes of
the vector $R$ may then be summarized as
\beq
\Gamma_{R}=\Gamma^{\chi\bar{\chi}} +  \sum_q \Gamma^{q\bar{q}}+\sum_\ell \Gamma^{\ell\bar{\ell}}+\sum_\nu  \Gamma^{\nu\bar{\nu}}+\Gamma^{WW}+\Gamma^{ZZ} +\Gamma^{\gamma Z}+ \Gamma^{ZH} + \Gamma^{X} \; ,
\eeq
where the formulae for the partial widths are provided in
Appendix~\ref{sec:widths}. For consistency of our description, we will
impose $\Gamma_{R}/m_R < 1$ which already gives a bound on all
coupling constants. If we define $g_\psi \equiv
\sqrt{(g^{V}_{\psi R})^2+(g^{A}_{\psi R})^2}$ for any fermion $\psi$ we then have the following
constraints on the couplings in isolation
\beq
g_{\phi R} \lesssim 12 \ ,  \quad g_\chi \lesssim 6 \ , \quad g_{\ell},g_{\nu} \lesssim 3.5  \ , \quad g_q \lesssim 1.5 \;.
\eeq
Here we assumed family-independent SM couplings: $g_u = g_d \equiv g_q$.

\subsection{Production and decay of $R$ at colliders}
\label{sec:production}

At colliders, the new spin-1 state $R$ can e.g.\ be produced via
Drell-Yan (DY) production, vector boson fusion (VBF) or `$R$-Strahlung' from a SM gauge boson. We will focus
in this paper on DY production. 
However, VBF will be important if the coupling of $R$ to
$W$'s is large, e.g.\ if $R$ arises from a composite theory. In this
case, there would be more search channels with 2 additional jets in
the final state.

In the case of DY production, we can decompose the cross-section for
the production of $R$ in association with an additional particle $Y$
and subsequent decay of $R$ into $xy$, as
\beq
\sigma(q \bar{q} \to R + Y \to xy + Y)= \sigma(q \bar{q} \to R+ Y) \cdot {\rm BR}(R \to xy) \;,
\eeq
where we have used the narrow width approximation (NWA), applicable if
$\Gamma_R/ m_R \ll 1$.

As a consequence of the NWA, the DY cross-section $\sigma(q\bar{q} \to
R \to xy)$ can be written as~\cite{Carena:2004xs,Accomando:2010fz}
\beq
\sigma_{xy} \propto \left[g_u^{2} w_u(s,m_{R}^2)+g_d^{2} w_d(s,m_{R}^2)\right] \cdot {\rm BR}({R \to xy}) \; ,
\label{eq:ll}
\eeq
where $w_{u,d}$ parameterise the parton distribution functions (PDFs) of the proton.
For a narrow resonance the only dependence of the $w_{u,d}$
coefficient on the resonance $R$ is through $m_{R}$.  Since the
functions $w_{u,d}$ are known, one can always translate a bound on
$g_u$ into a bound on $g_d$ or a bound on a model with a given ratio
of $g_u / g_d$ into a bound on a different model. For this reason, we
will show experimental bounds assuming $g_u = g_d \equiv
g_q$ for simplicity.

\begin{figure}[tb]
\begin{center}
{
\centering
\includegraphics[width=0.55\columnwidth]{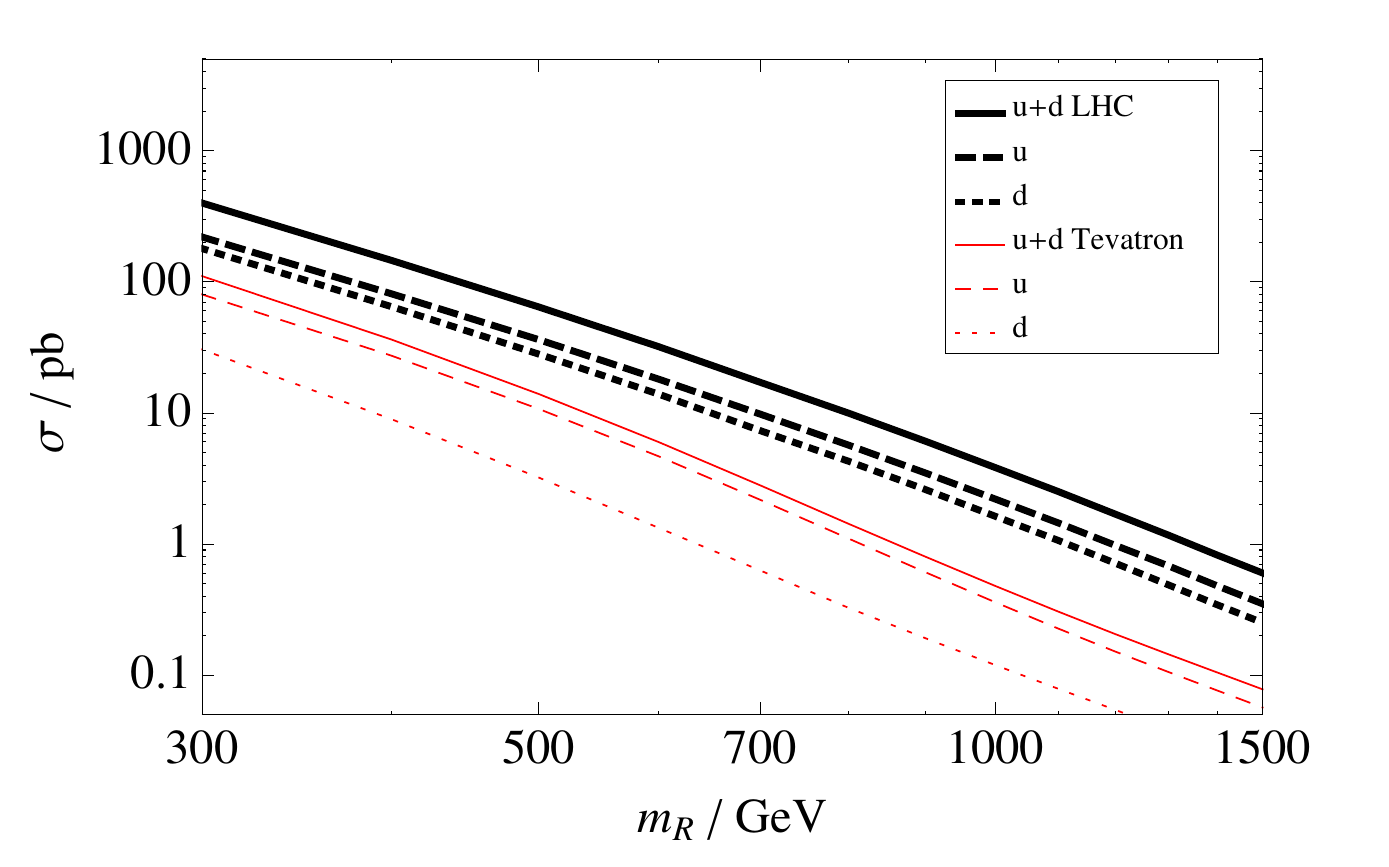}
}
\end{center}
\vspace*{-3mm}
\caption{Drell-Yan production of $R$ as a function of $m_R$ at
  the LHC (with $\sqrt{s}=7$ TeV) and at the Tevatron for
  $g_{u,d}^{A,V}$ equal to those of the SM $Z$ boson (solid lines). We also show the production cross-section for the case that $R$ couples either only to $u$-quarks or only to $d$-quarks.}
\label{fig:prodcs}
\end{figure}

In Figure~\ref{fig:prodcs} we show the leading order (LO) cross-section for $\sigma(pp
\to R)$, with the couplings of $R$ to SM fields set equal to those of
the SM $Z$ boson as well as the individual contributions from $u$-quarks and $d$-quarks. The production cross-section of $R$ with different
couplings may be found by a simple rescaling.  Note that QCD
corrections can enhance the DY production significantly; in the following,
we take the K-factors from~\cite{Accomando:2010fz} for the MSTW08 NNLO
PDF.

Once we know the production cross-section of $R$ for given couplings $g_q$, we can translate LHC bounds on the cross-section for a certain final state $xy$ into a bound on the product $g_q^2 \cdot \text{BR}(R \rightarrow xy)$. The decay modes of $R$ into the $X$ sector can appear at colliders as
additional missing energy, displaced vertices or high-multiplicity SM
final states.  An example is the decay mode $R \to Z h'$ where $h'$ is
a new scalar state responsible for the mass of $R$.  Another contribution to $\Gamma^X$ could come from decay
modes of $R$ to additional hidden sector states. A number of such
possibilities were considered e.g.~in~\cite{Strassler:2006im}.

\section{Collider bounds}
\label{sec:bounds}

\begin{table}[!h]
\setlength{\tabcolsep}{5pt}
\renewcommand{\arraystretch}{1.7}
\center
\begin{tabular}{||c|c|c|c|c||} 
\hline \hline 
Channel [Exp] &  L  $[{\rm fb}^{-1}]$ & Mass range & Couplings & Reference
\\
\hline
$pp \rightarrow  j + \slashed{p}_\mathrm{T} \,\, [{\rm ATLAS}] $ & 1.0 & $*$ & $g_q \, g_\chi$ &\cite{Aad:2011wc}   
\\
$pp \rightarrow  j + \slashed{p}_\mathrm{T} \,\, [{\rm CMS}] $ & 4.7 & $*$ & $g_q\, g_\chi$ &\cite{CMS-monojet}   
\\
$pp \rightarrow  j + \slashed{p}_\mathrm{T} \,\, [{\rm CDF}] $ & 6.7 & $*$ & $g_q \,g_\chi$ &\cite{CDF-monojet}   
\\
\hline
$pp \rightarrow  \gamma + \slashed{p}_\mathrm{T} \,\, [{\rm CMS}] $ & 1.14 & $ * $ & $g_q \, g_\chi \, , \ g_q\, g^R_{Z\gamma}$ &\cite{CMS-Monophoton1}   
\\
$pp \rightarrow  \gamma + \slashed{p}_\mathrm{T} \,\, [{\rm CMS}] $ & 4.7 & $ * $ & $g_q \, g_\chi \, , \ g_q\, g^R_{Z\gamma}$ &\cite{CMS-Monophoton2}   
\\
\hline
$pp \rightarrow  j \,j \,\, [{\rm ATLAS}] $ & 1.04 & $900-4000$ & $g_q $ &\cite{Aad:2011fq}   
\\
$pp \rightarrow  j \,j \,\, [{\rm CDF}] $ & 1.13 & $250-1400$ & $g_q $ &\cite{Aaltonen:2008dn}   
\\
\hline
$pp \rightarrow  \ell  \ell  \,\, [{\rm CMS}] $ & 4.9 & $300-2500$ & $g_q\, g_\ell$ &\cite{CMS-Dileptons}
\\
$pp \rightarrow  \ell  \ell  \,\, [{\rm ATLAS}] $ & 1.08--1.21 & $200-2000$ & $g_q\, g_\ell$ &\cite{Collaboration:2011dca}
\\
$pp \rightarrow  \tau \tau  \,\, [{\rm CMS}] $ & 4.9 & $350-1600$ & $g_q\, g_\tau$ &\cite{CMS-ditaus}
\\
\hline
$pp \rightarrow  Z \, Z^{(*)} \,\, [{\rm ATLAS}] $ & 4.9 & $110-600$ & $g_q\, g^R_{ZZ}$ &\cite{ATLAS-Higgs}
\\
$pp \rightarrow  Z \, Z \,\, [{\rm ATLAS}] $ & 1.0 & $320-1500$ & $g_q\, g^R_{ZZ}$ &\cite{ATLAS:2012iu}
\\
$pp \rightarrow  Z \, Z^{(*)} \,\, [{\rm CMS}] $ & $4.6-4.8$ & $110-600$ & $g_q\, g^R_{ZZ}$ &\cite{Chatrchyan:2012tx}   
\\
\hline
$pp \rightarrow  W \, W^{(*)} \,\, [{\rm ATLAS}] $ & 4.9& $110-600$ & $g_q\, g^R_{WW}$ &\cite{ATLAS-Higgs} 
\\
$pp \rightarrow  W \, W^{(*)} \,\, [{\rm CMS}] $ & $4.6-4.8$ & $110-600$ & $g_q\, g^R_{WW}$ &\cite{Chatrchyan:2012tx} 
\\
\hline
$pp \rightarrow  Z H \,\, [{\rm ATLAS}] $ & 4.7& $*$ & $g_q\, g^R_{ZH}$ &\cite{ATLAS-ZH} 
\\
\hline
$pp \rightarrow  t \, \bar{t} \text{ (boosted)}\,\, [{\rm CMS}] $ & 4.6& $1000-3000$  & $g_q\, g_t$&\cite{CMS-tt} 
\\
$pp \rightarrow  t \, \bar{t} \text{ (semileptonic)} \,\, [{\rm CMS}] $ & 4.9& $500-1500$  & $g_q\, g_t$&\cite{CMS-semitop} 
\\
\hline \hline
\end{tabular}
\caption{Collider searches in final states that constrain the
  couplings of $R$. Fields marked with a $*$ correspond to searches
  that do not look for heavy resonances and which, consequently, give
  constraints for arbitrary $m_R$. In the case of scalar DM $g_\chi$ should be replaced by $g_{\phi R}$.}
\label{table:darkZpsearches} \vspace{-0.35cm}
\end{table}

\begin{figure}[tb]
\begin{minipage}[b]{0.48\linewidth}
\centering
\includegraphics[width=.99\linewidth]{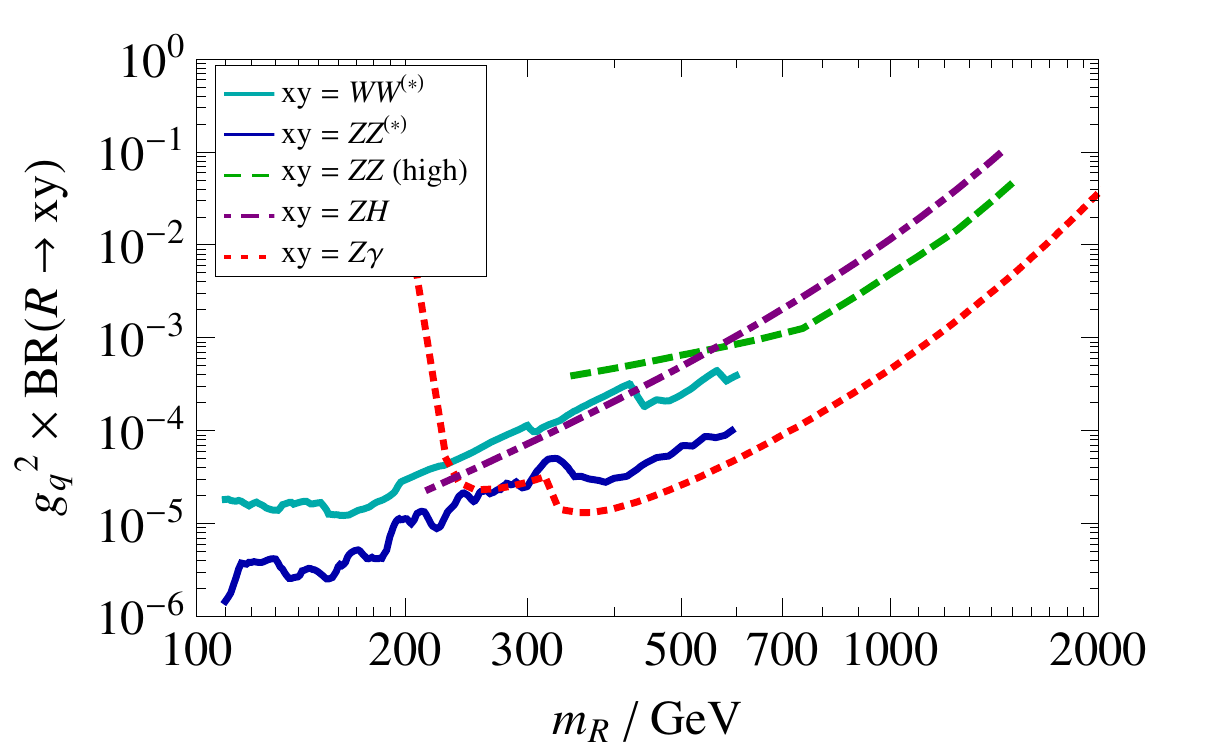}
\vspace*{-3mm}
\caption{Bounds on $g_q^2 \cdot \text{BR}(R \rightarrow xy)$ as a function of $m_R$, with
  $xy$ being either SM gauge bosons or $ZH$.}
\label{fig:bosons}
\end{minipage}
\hspace{0.5cm}
\begin{minipage}[b]{0.48\linewidth}
\centering
\includegraphics[width=.99\linewidth]{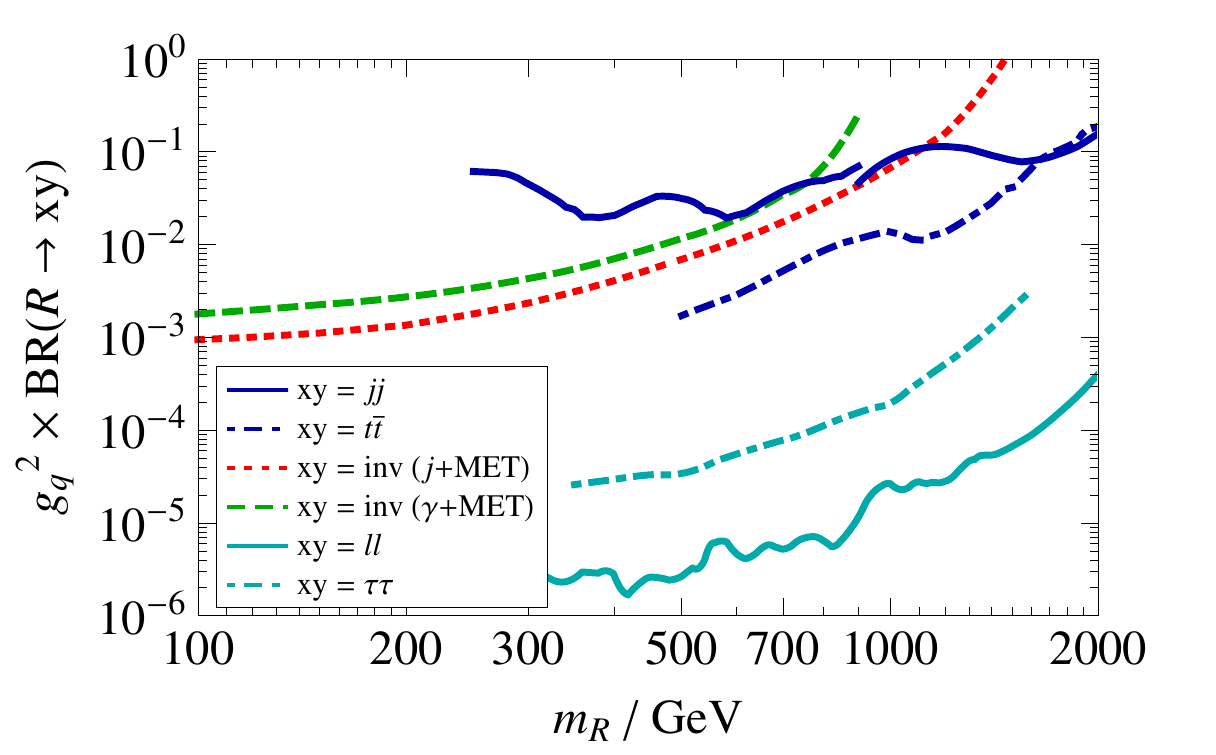}
\vspace*{-3mm}
\caption{Bounds on $g_q^2 \cdot \text{BR}(R \rightarrow xy)$ as a function of $m_R$, with
  $xy$ being either SM fermions or DM particles.}
\label{fig:fermions}
\end{minipage}
\end{figure}

In this section we summarize the collider bounds from LHC and Tevatron for the decay modes of $R$ and then discuss each decay mode in more detail.
Table~\ref{table:darkZpsearches} lists the current collider searches
we consider here, while the corresponding limits we derive are
summarised in Figure~\ref{fig:bosons} and Figure~\ref{fig:fermions}. The confidence level for all bounds is at least $95\%$.

\subsection{Monojet searches}

The monojet final state arises from production of $R$ with an additional jet, $j=q,g$, followed by the decay of $R$ into DM
or neutrinos: $\sigma(pp\to j R \to j \, \slashed{p}_\mathrm{T}
)$. Decay modes into additional hidden sector states also contribute
to the monojet signal, provided these states do not decay back to SM
states within the detector.  

Monojet searches have been performed at the
Tevatron~\cite{CDF-monojet, Abazov:2003gp}, and at the LHC by both
CMS~\cite{CMS-monojet} and ATLAS~\cite{ATLAS-monojet} with similar
sensitivity. To calculate our bounds we compare the limits from ATLAS
with the parton-level monojet signal from $R$ simulated using {\tt
  CalcHEP}~\cite{Pukhov:2004ca}. 
For monojet searches, the jets have sufficiently high $p_\mathrm{T}$ that the errors from neglecting
parton showering and hadronization are small (see
e.g.~\cite{Bai:2010hh,Choudalakis:2011bf}).

\begin{figure}[tb]
\begin{minipage}[t]{0.48\linewidth}
\centering
{\includegraphics[height=0.64\linewidth]{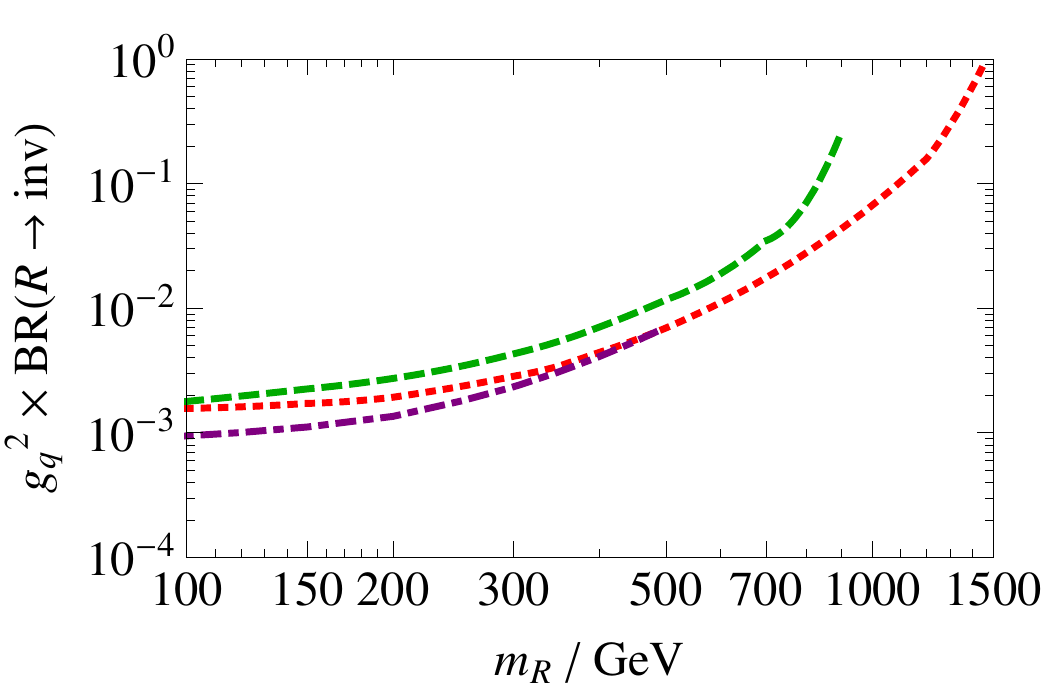}
}
\vspace*{-3mm}
\caption{Limits on $g_q^2 \cdot \text{BR}(R \rightarrow \text{inv})$
  from 
  monojet searches (red, dotted
  and purple, dot-dashed)~\cite{Aad:2011wc} and the monophoton
  searches (green, dashed)~\cite{CMS-Monophoton2}. The red dotted line
  corresponds to the cut $p_\mathrm{T} > 350$~GeV, the purple
  dot-dashed line to $p_\mathrm{T} > 250$~GeV.  }
\label{fig:monophoton}
\end{minipage}
\hspace{0.5cm}
\begin{minipage}[t]{0.48\linewidth}
{\includegraphics[height=0.64\linewidth]{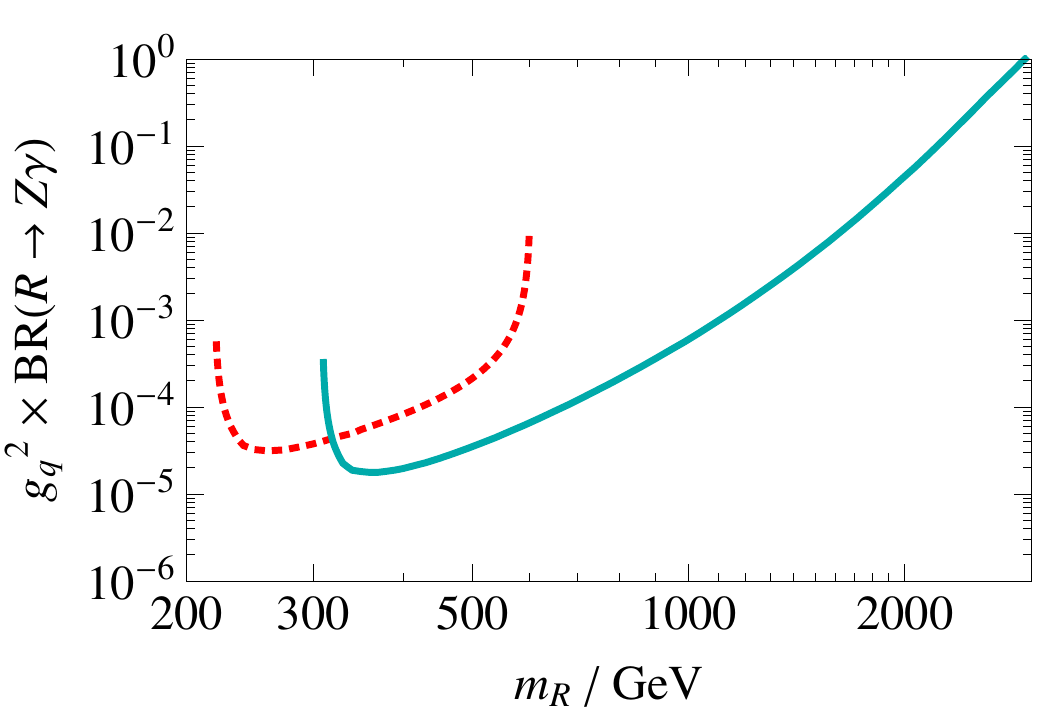}
}
\vspace*{2.2mm}
\caption{Limits on $g_q^2 \cdot \text{BR}(R \rightarrow Z\gamma)$ from
  the monophoton searches in Ref.~\cite{CMS-Monophoton1} (red, dotted)
  and Ref.~\cite{CMS-Monophoton2} (light blue, solid).  }
\label{fig:Zgamma}
\end{minipage}
\end{figure}

For $m_R \gtrsim 400$~GeV, we find that the ATLAS search with $p_\mathrm{T}(j) > 350$~GeV gives the strongest constraint.
For lighter $R$, a stronger bound is obtained from the ATLAS search with $p_\mathrm{T}(j) > 250$~GeV. 
The Tevatron gives the strongest bound for $m_R \lesssim 100$~GeV because of its high luminosity and lower monojet $p_\mathrm{T}$ cut.
\mbox{Our results are shown in Figure~\ref{fig:monophoton}.} 

\subsection{Monophoton searches}

The monophoton final state can arise from two different processes.
The first is similar to the monojet process with production of $R$
and initial state radiation of a photon, followed by decay of $R$ into
DM or neutrinos $\sigma(pp\to \gamma R \to \gamma \, \slashed{p}_\mathrm{T} )$. The second possibility is DY production of $R$
followed by the direct decay of $R$ into $Z \gamma$, with subsequent
decay of $Z$ to neutrinos $\sigma(pp\to R \to \gamma Z \to \gamma \,
\nu \bar{\nu} )$.

We find that the initial state radiation of a photon provides weaker
constraints on the invisible branching ratio of $R$ than the initial
state radiation of a jet (see Figure~\ref{fig:monophoton}). The second
process, however, offers an interesting possibility to limit the
branching ratio of $R \rightarrow Z\gamma$. Note, however, that such
monophoton searches are only sensitive to these decays if $m_R/2$ is
sufficiently larger than any cut on 
$\slashed{p}_\mathrm{T}$ or the photon $p_\mathrm{T}$. We use the data
from~\cite{CMS-Monophoton1,CMS-Monophoton2} and obtain the limit
curves shown in Figure~\ref{fig:Zgamma} using {\tt CalcHEP}. Note that the limits on the
direct decay $R\to Z \gamma$ have been calculated without including
the contribution of invisible decays of $R$ together with a photon
(from initial state radiation). For sizeable invisible branching of $R$,
the bounds would become even stronger.

\subsection{Dijet resonances}

Searches for resonances in the invariant mass distribution of dijet
events have been carried out at Tevatron and at the LHC. We use the
CDF limit~\cite{Aaltonen:2008dn} for $m_{R} < 900$~GeV and the
ATLAS limits~\cite{Aad:2011fq} for $m_{R} \geq 900$~GeV. In the latter
case, the bound on the cross-section is not quite independent of the
width of the resonance, which depends on $\Gamma_R$ and on the
detector resolution (see also~\cite{Choudalakis:2011bf}).

\begin{figure}[tb]
\begin{minipage}[t]{0.48\linewidth}
\centering
\includegraphics[height=.64\linewidth]{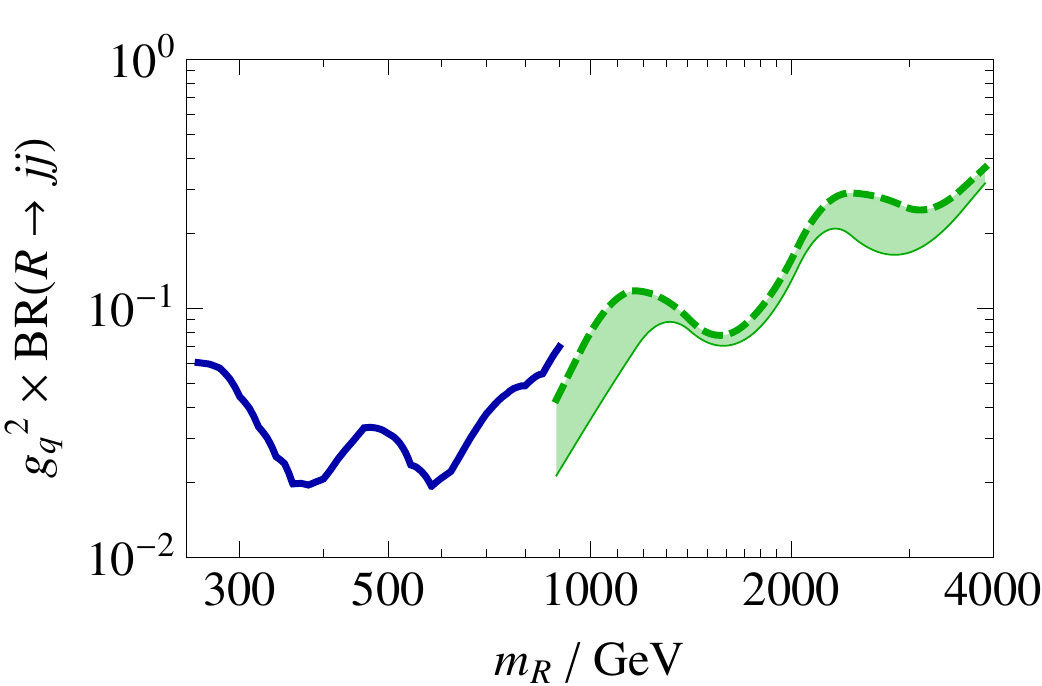}
\vspace*{2.2mm}
\caption{Combined dijet limits from CDF (blue, solid) and ATLAS
  (green, dashed).The line width reflects the dependence of the ATLAS bound
  on $\Gamma_{R}$ (which is varied between 10\% and 25\% of the
  mediator mass). 
}
\label{fig:dijet-combined}
\end{minipage}
\hspace{0.5cm}
\begin{minipage}[t]{0.48\linewidth}
\centering
{\includegraphics[height=0.64\linewidth]{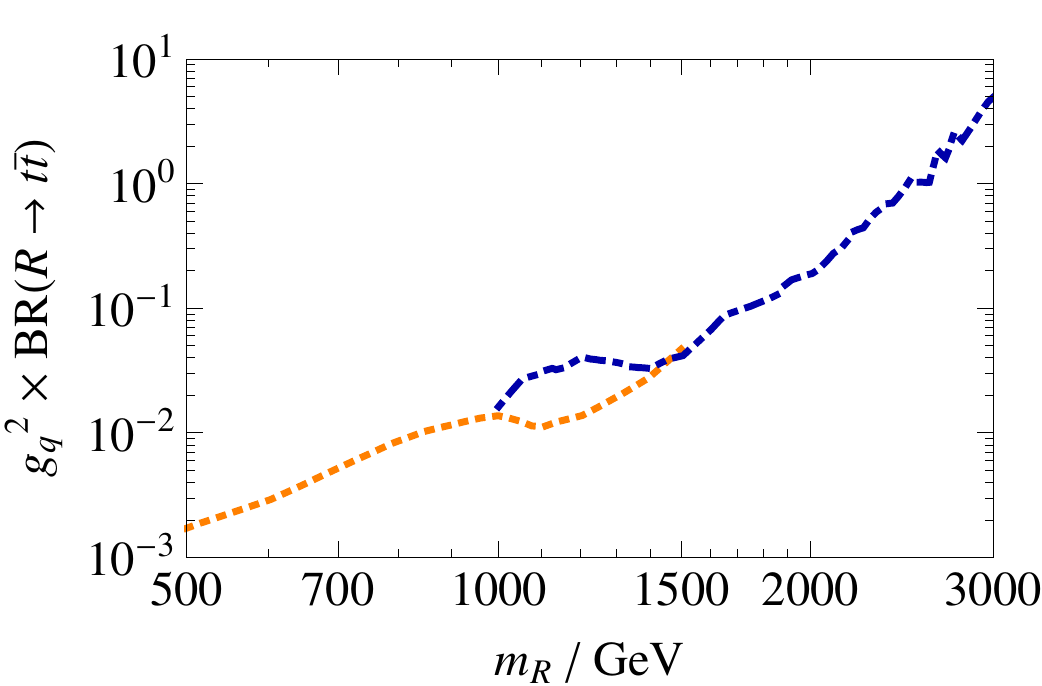}
}
\vspace*{-3mm}
\caption{Limit on $g_q^2 \cdot \text{BR}(R \rightarrow t {\bar t})$
  from searches for boosted tops (blue, dot-dashed)~\cite{CMS-tt}
  and semileptonic tops (orange, dotted)~\cite{CMS-semitop}.}
\label{fig:tops}
\end{minipage}
\end{figure}

To estimate this dependence we have generated $m_{jj}$ distributions
for different values of $\Gamma_R$ using {\tt LanHEP}~\cite{Semenov:2010qt}
and {\tt CalcHEP} and convoluted these distributions with the detector
resolution~\cite{Aad:2011fq}. By comparing the resulting
width of the peak to the bounds given in Table II of~\cite{Aad:2011fq}
we estimate how the limit varies with the mediator width for
$\Gamma_R/m_R$ in the range 0.1-- 0.25. The result is shown in
Figure~\ref{fig:dijet-combined}
together with the limits from dijet searches from CDF. 
We take the upper end of the band shown
in Figure~\ref{fig:dijet-combined} and conservatively apply it as a bound for all
widths $\Gamma_R/m_R < 0.25$.

\subsection{Top pairs}

Searches for dijet resonances constrain the decays of $R$ into the
five lightest quarks. To constrain $g_q^2 \cdot \text{BR}(R
\rightarrow t\bar{t})$ independently, we use the dedicated CMS
searches~\cite{CMS-tt,CMS-semitop} for $t {\bar t}$ resonances. The
resulting bounds are shown in Figure~\ref{fig:tops}.

We note that for family independent couplings and assuming $m_R > 2
m_t$, we can always use a bound on $g_q^2 \cdot \text{BR}(R
\rightarrow jj)$ to infer a bound on $g_q^2 \cdot \text{BR}(R
\rightarrow t\bar{t})$ using the relation
\begin{equation}
\text{BR}(R \rightarrow t\bar{t}) = \frac{\sqrt{1 - 4 m_t^2 / m_R^2}}{2 + 3 g_d^2/g_u^2} \cdot \text{BR}(R \rightarrow jj)\; .
\end{equation}
 For $g_u \sim g_d$, the bound on
$g_q^2 \cdot \text{BR}(R \rightarrow t\bar{t})$ inferred from the
dijet limit is comparable to the direct bound from top pair
searches. As expected, for $g_u \ll g_d$, the inferred bound on $g_q^2
\cdot \text{BR}(R \rightarrow t\bar{t})$ is much stronger than the
direct one, while for $g_u \gg g_d$, we can actually invert the equation
above to obtain a bound on $g_q^2 \cdot \text{BR}(R \rightarrow jj)$
from the bounds on $g_q^2 \cdot \text{BR}(R \rightarrow t {\bar t})$.

\begin{figure}[tb]
\begin{minipage}[t]{0.48\linewidth}
\centering
{\includegraphics[height=0.64\linewidth]{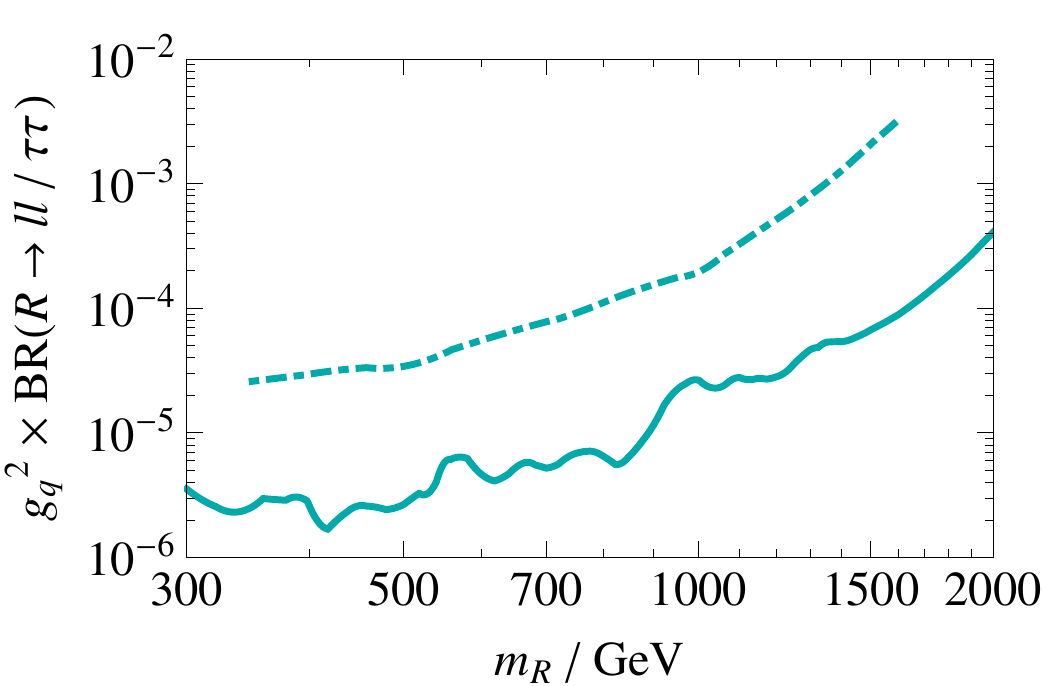}
}
\vspace*{-3mm}
\caption{Limit on $g_q^2 \cdot \text{BR}(R \rightarrow \ell\ell)$ (solid) from the
  CMS search for dilepton resonances~\cite{CMS-Dileptons} and on $g_q^2 \cdot \text{BR}(R \rightarrow \tau\tau)$ (dotted)
  from the corresponding ditau search~\cite{CMS-ditaus}.}
\vspace{26pt}
\label{fig:dilepton}
\end{minipage}
\hspace{0.5cm}
\begin{minipage}[t]{0.48\linewidth}
\centering
{\includegraphics[height=0.64\linewidth]{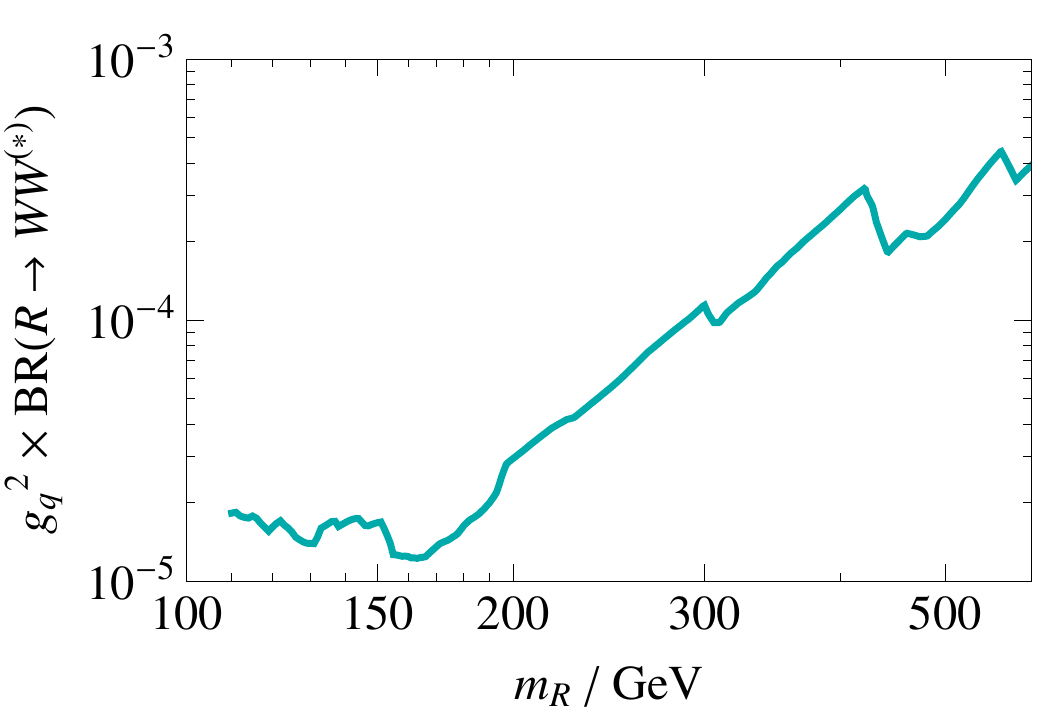}
}
\vspace*{2.2mm}
\caption{Limit on $g_q^2 \cdot \text{BR}(R \rightarrow WW)$ from
  combined searches for Higgs decays to $WW$~\cite{ATLAS-Higgs}.}
\vspace{26pt}
\label{fig:WW}
\end{minipage}
\end{figure}

\subsection{Dilepton resonances}

We consider the recent CMS search for dilepton
resonances~\cite{CMS-Dileptons} with $\ell = (e,\; \mu)$ as well as the search for ditau resonances~\cite{CMS-ditaus} and show the resulting limits in
Figure~\ref{fig:dilepton}. 
If we assume family independent couplings, we can obtain a stronger bound on the ditau channel making use of the relation $\text{BR}(\tau \tau) = \text{BR}(\ell\ell)/2$.

\subsection{$WW$ and $ZZ$}

\begin{figure}[tb]
\begin{minipage}[t]{0.48\linewidth}
\centering
{\includegraphics[height=0.64\linewidth]{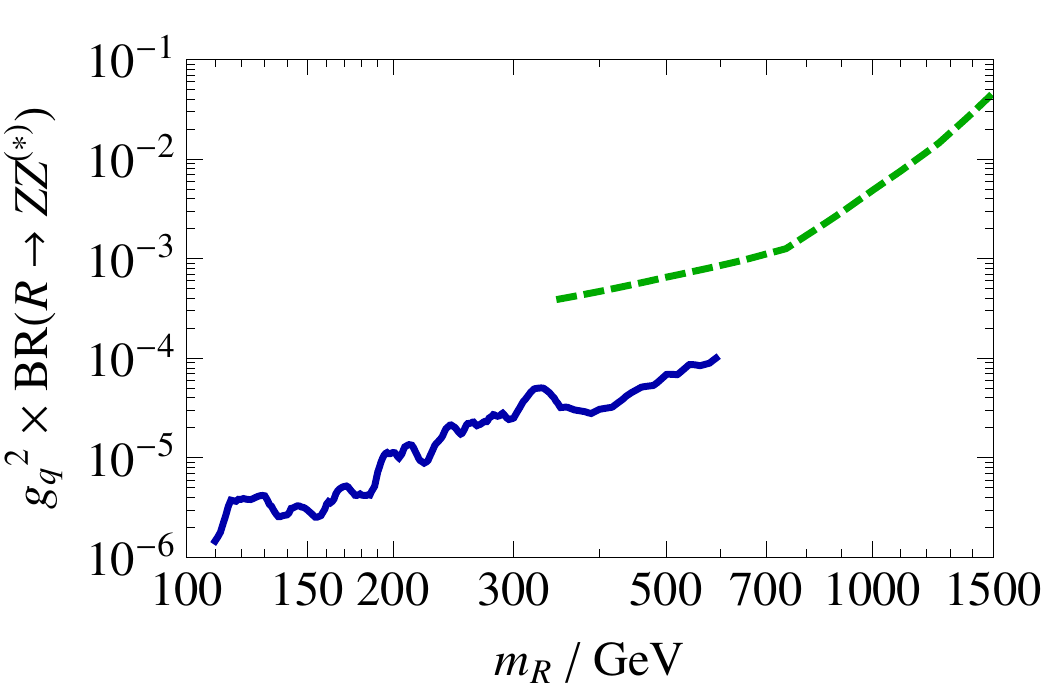}
}
\vspace*{-3mm}
\caption{Limit on $g_q^2 \cdot \text{BR}(R \rightarrow ZZ)$ from
  combined searches for Higgs decays to $ZZ$~\cite{ATLAS-Higgs} (blue,
  solid) and from a dedicated search for high-mass $ZZ$
  resonances~\cite{ATLAS:2012iu} (green, dashed).}
\label{fig:ZZ}
\end{minipage}
\hspace{0.5cm}
\begin{minipage}[t]{0.48\linewidth}
\centering
{\includegraphics[height=0.64\linewidth]{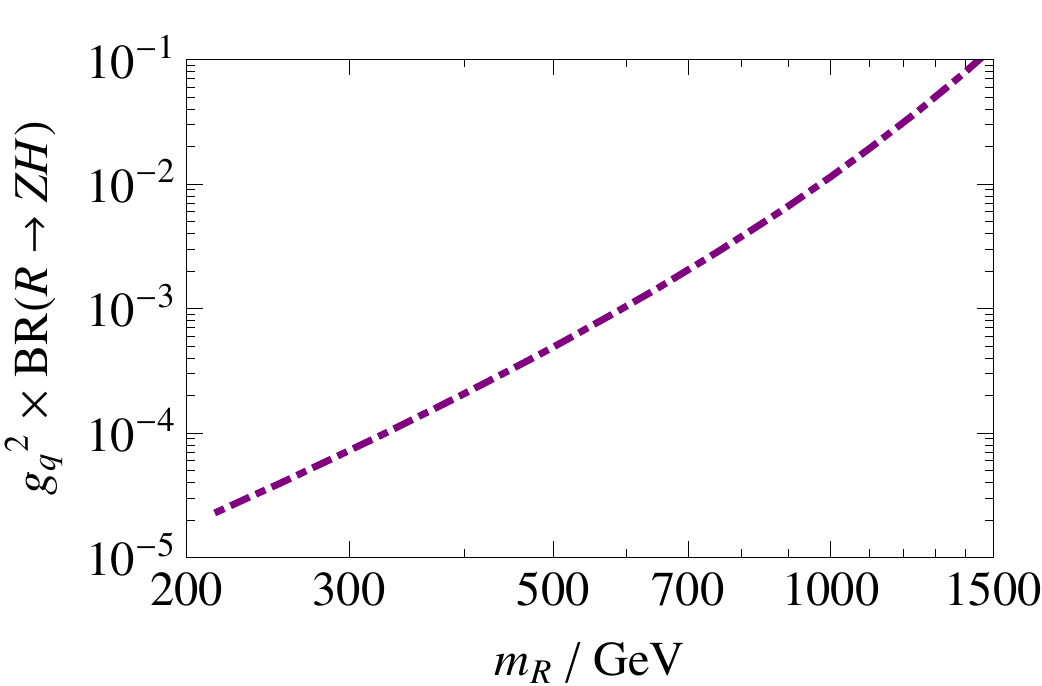}
}
\vspace*{-3mm}
\caption{Limit on $g_q^2 \cdot \text{BR}(R \rightarrow ZH)$ from the
  ATLAS search for associate production of Higgs and $Z$ for a Higgs
  mass of \mbox{$m_H = 125$~GeV}~\cite{ATLAS-ZH}.}
\label{fig:ZH}
\end{minipage}
\end{figure}

Searches for $WW$ and $ZZ$ final state have been performed in
the context of Higgs searches at the
LHC~\cite{ATLAS-Higgs,Chatrchyan:2012tx} and as a dedicated search for
high-mass resonances~\cite{ATLAS:2012iu}. The results of these
searches are shown in Figures~\ref{fig:WW} and~\ref{fig:ZZ}. We observe
that the Higgs search limits from $WW$ and $ZZ$ on the $R$
couplings are significant even for relatively large $m_R$. The reason
is that for a vector boson, the coupling involves a derivative, so it
is enhanced compared to e.g.\ the SM Higgs coupling for large masses.

\subsection{$ZH$}
Recent results from the LHC exclude the SM Higgs with a mass between
130 and 550~GeV at the 99\% confidence level
\cite{Chatrchyan:2012tx,ATLAS-Higgs}, while the combined LEP2 results
exclude it below 114.5~GeV at the 95\% confidence level
\cite{Barate:2003sz}. In the allowed mass range, we have upper limits
on the cross-section $pp \rightarrow ZH$ from searches for Higgs
production in association with a SM vector. We take the limit from $pp
\rightarrow ZH \rightarrow \nu \bar{\nu} b \bar{b}$
from~\cite{ATLAS-ZH}, giving $\sigma_{ZH} < 1.7 \mathellipsis 2.0$ pb
depending on the value of the Higgs mass in the allowed range. We show
the resulting limit for $m_H = 125$~GeV in Figure~\ref{fig:ZH}.

\section{Implications for dark matter direct detection}
\label{sec:directdetection}

In this section, we apply the collider bounds obtained above to constrain
DM-nucleon interactions mediated by $R$. We can
calculate the direct detection cross-section by integrating out both
$R$ and $Z$ to generate the corresponding effective operators.
In the case of Dirac DM we obtain
\begin{align}
  {\cal L}_{\chi}^\mathrm{eff} = & \, b^\mathrm{V}_{f} \bar{\chi} \gamma_\mu
  \chi \bar{f} \gamma^\mu f\, + \, b^\mathrm{A}_{f} \bar{\chi} \gamma_\mu
  \gamma^5 \chi \bar{f} \gamma^\mu
  \gamma^5 f  \; ,
\label{eq:bf}
\end{align}
where we have neglected terms that vanish in the non-relativistic
limit and defined the effective couplings 
\begin{align}
  b^\mathrm{A,V}_{f} = b^\mathrm{A,V}_{f R} + b^\mathrm{A,V}_{f Z} =
  \frac{g^\mathrm{A,V}_{\chi R} g^\mathrm{A,V}_{f R}}{m_R^2} +
  \frac{g^\mathrm{A,V}_{\chi Z} g^\mathrm{A,V}_{f Z}}{m_Z^2} \; .
\end{align}

Unless $b^\mathrm{V}$ is very small compared to $b^\mathrm{A}$, the
direct detection cross-section will be dominated by the effective
vector-vector interaction between the DM particle and nucleons $(p,n)$
given by
\begin{align}
  {\cal L}^\mathrm{V}_\chi = \, f_{p}^\chi \bar{\chi} \gamma_\mu \chi
  \bar{p}\gamma^\mu p + f_{n}^\chi \bar{\chi} \gamma_\mu \chi \bar{n}
  \gamma^\mu n\, ; \quad
  f_p^\chi = 2 b^\mathrm{V}_{u} + b^\mathrm{V}_{d} \ , \ f_n^\chi = 2
  b^\mathrm{V}_{d} + b^\mathrm{V}_{u} \, .
  \label{eq:fnfp}
\end{align}

In the case of the complex scalar DM we have, similarly,
\begin{align}
  {\cal L}_{\phi}^\mathrm{eff} = & \, a^\mathrm{V}_{f} \ J^\mu_\phi \bar{f} \gamma_\mu f\ , \ 
   {\cal L}^\mathrm{V}_\phi = \, f_{p}^\phi J^\mu_\phi
  \bar{p}\gamma_\mu p + f_{n}^\phi J^\mu_\phi \bar{n}
  \gamma_\mu n\, , \quad  
\end{align}
where 
\begin{align}
  a^\mathrm{V}_{f} = a^\mathrm{V}_{f R} + a^\mathrm{V}_{f Z} =
  \frac{g_{\phi R} g^\mathrm{V}_{f R}}{m_R^2} +
  \frac{g_{\phi Z} g^\mathrm{V}_{f Z}}{m_Z^2} \; ; \ \
   f_p^\phi = 2 a^\mathrm{V}_{u} + a^\mathrm{V}_{d} \ , \ f_n^\phi = 2
  a^\mathrm{V}_{d} + a^\mathrm{V}_{u} \, .
\end{align}

Because of the conservation of the vector current, there is no
contribution of sea quarks or gluons to the effective couplings. 
For both Dirac and complex scalar DM we obtain the DM-nucleon cross-section 
\begin{equation}
\sigma_N = \mu^2_{\chi N} f_N^2 /\pi \;, \text{ where } \ N=p,\,n \; . 
\label{eq:DD}
\end{equation}

In the following, we will consider two different possibilities for
generating effective interactions of 
nucleons
and DM particles. First we
consider the case where $R$ has sizeable direct couplings to
quarks and all other couplings are arbitrary. 
Afterwards
we consider the
case where the interaction state $X$ corresponding to the mass eigenstate 
$R$ couples only to the DM particle
and couplings to SM particles are generated only via mixing. An
example would be the `dark' $Z'$, where $R$ is the gauge boson of a
new $U(1)$ under which only the DM particle is charged. In this case,
we can use collider bounds to directly constrain the mixing
parameters, and therefore the direct detection cross-section.

\subsection{Direct detection through direct couplings}

Let us start with the general case where $R$ can have arbitrary
couplings to SM particles. The only assumption we make is that $R$ has
a sizeable branching into quarks.
This assumption is important for two reasons. First, we want to
exclude the case where $g_q$ is so small that the total number of
$R$-particles produced at the LHC is insufficient to give a detectable
monojet signal. We will come back to the case where
$g_{\chi} \gg g_{u R}^V , \; g_{d R}^V$ in
Section~\ref{sec:ddmixing}.

Second, this assumption ensures that DM direct detection is dominated
by $R$-exchange, with $Z$-exchange giving only a negligible
contribution. This case is interesting because it
allows a ratio $f_n / f_p$ significantly different from
the one for $Z$ exchange. In fact defining $y \equiv g_{uR}^V / g_{dR}^V$ we obtain $f_n
/ f_p = (y + 2)/(2y + 1)$, which can in principle take any arbitrary
value.

We then obtain from Equation~(\ref{eq:DD})
\beq
\sigma_p  \simeq (2y+1)^2 \frac{ \mu_{\chi n}^2}{\pi} \frac{\left(g_{d R}^V\right)^2 \left(g_{\chi R}^V\right)^2}{m_R^4} \leq (2y+1)^2 \frac{ \mu_{\chi n}^2}{\pi} \frac{g_d^2 g_\chi^2}{m_R^4} \;.
\eeq
For $m_\chi \ll m_R$ we can use Equation~(\ref{eq:widthchi}) to obtain 
\beq
\sigma_p \leq 12 (2y+1)^2 \frac{\mu_{\chi n}^2 \Gamma_R}{m_R^5} g_d^2 \cdot \text{BR}(R \rightarrow \text{inv}) \; .
\eeq
As discussed in Section~\ref{sec:bounds}, monojet and monophoton searches
at the LHC provide a limit on $g_d^2 \cdot \text{BR}(R \rightarrow
\text{inv})$, so that we obtain a bound on the direct detection
cross-section if we can constrain $\Gamma_R$.

\begin{figure}[tb]
\begin{minipage}[b]{0.48\linewidth}
\centering
{\includegraphics[width=0.95\linewidth]{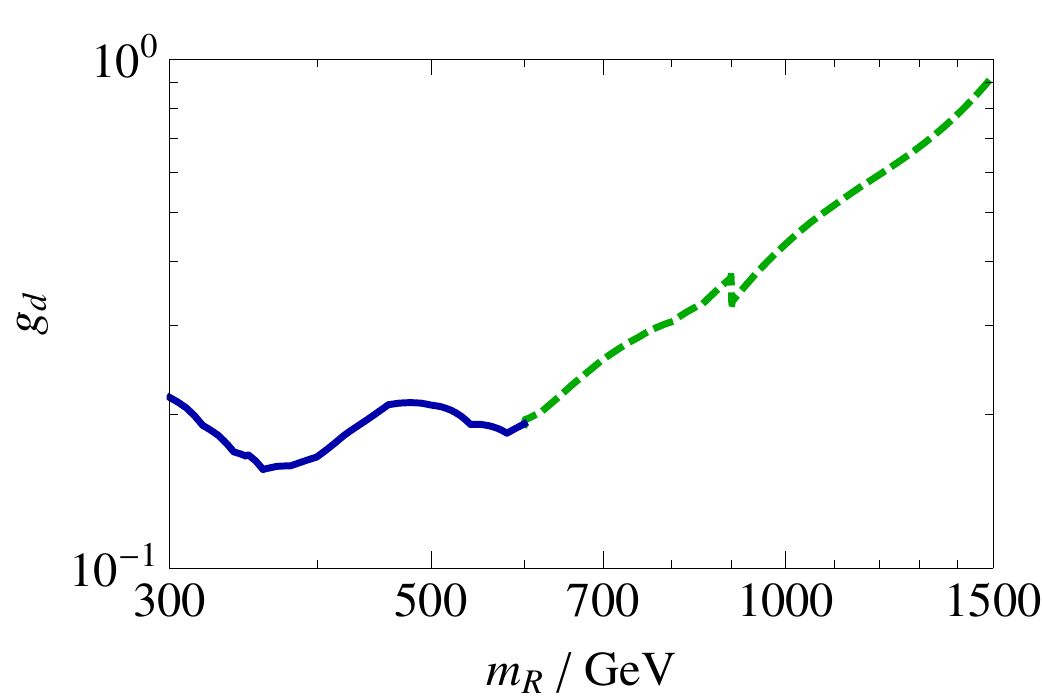}
}
\vspace*{-3mm}
\caption{Limit on $g_d$ from the combination of all experimental
  bounds (see text).
}
\vspace*{0mm}
\label{fig:gd}
\end{minipage}
\hspace{0.5cm}
\begin{minipage}[b]{0.48\linewidth}
\centering
{\includegraphics[width=0.95\linewidth]{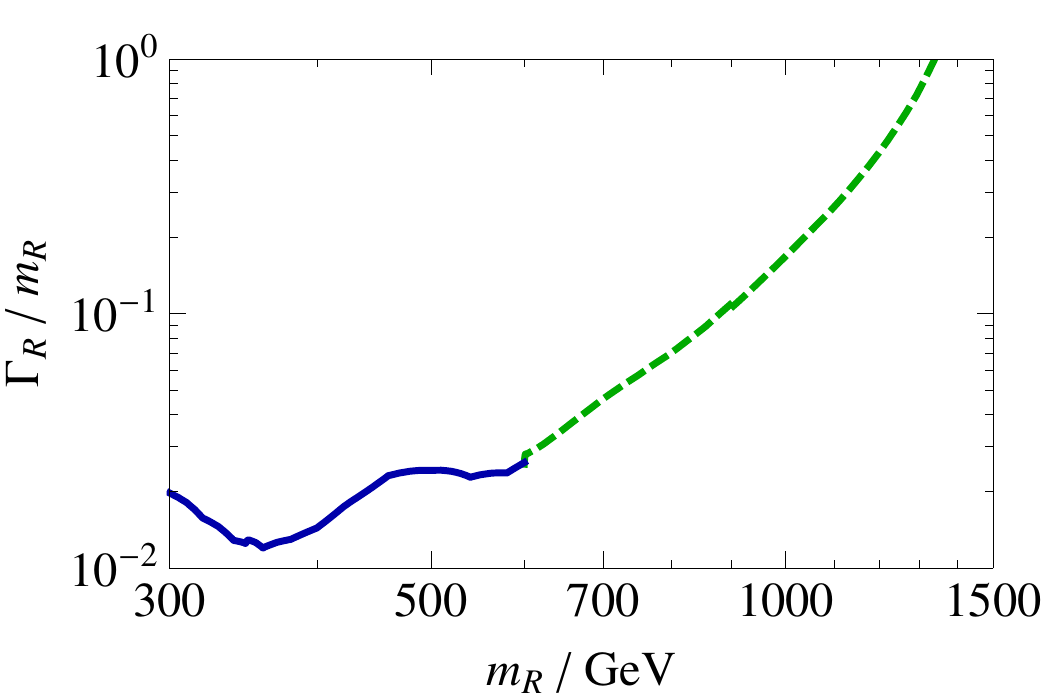}
}
\vspace*{-3mm}
\caption{Limit on $\Gamma_R$ from the combination of all
  experimental bounds.}
\label{fig:Gamma}
\end{minipage}
\end{figure}

Of course, if we allow decays into new states that give complicated
experimental signatures, we can make $\Gamma_R$ arbitrarily
large. Therefore, we will now assume that all new states are either SM
particles, or remain invisible, i.e.~escape the detector without
decaying into visible particles. In that case, we can combine the
bounds from Section~\ref{sec:bounds} to obtain an upper limit on $g_d$
and, assuming family independent couplings, constrain
$\Gamma_R$. The resulting bounds are shown in
Figures~\ref{fig:gd} and~\ref{fig:Gamma}.

With present data, we can  constrain $g_d$ and $\Gamma_R$ only
in the range $300\text{ GeV} \leq m_R \leq 600 \gev$. However, the
only decay channel that is presently not available above 600~GeV is $R
\rightarrow WW$. We simply assume that upcoming searches for
this decay mode will give bounds comparable to the current bounds for
$R \rightarrow ZZ$. Consequently, we assume that the bound on $g_q^2
\cdot \text{BR}(R \rightarrow ZZ)$ also applies to $g_q^2 \cdot
\text{BR}(R \rightarrow WW)$ so that we can extend our
analysis up to 1200~GeV.  Even a somewhat weaker bound on $g_q^2 \cdot
\text{BR}(R \rightarrow WW)$ would not change our results
dramatically, because decays into $WW$  give only a subdominant
contribution to the total width of $R$.

\begin{figure}[t]
\begin{center}
{
\includegraphics[height=.30\columnwidth]{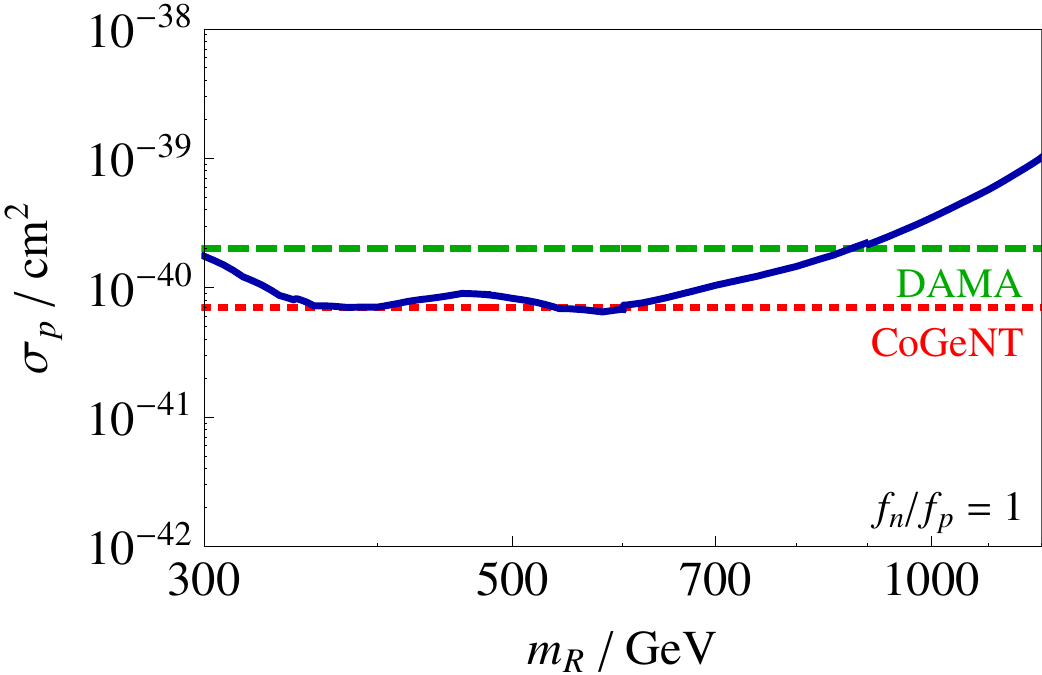}
\includegraphics[height=.312\columnwidth]{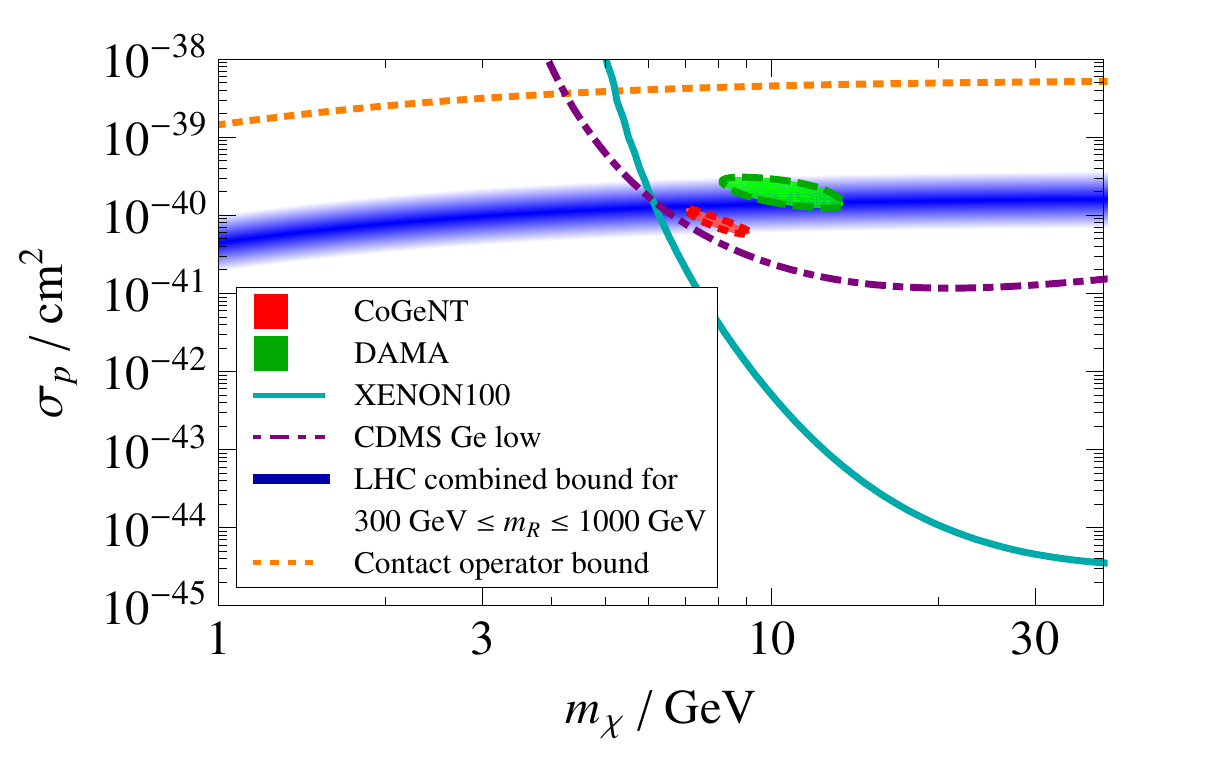}
\vspace*{-3mm}
\caption{Left: Bound from LHC data on the direct detection
  cross-section as a function of the mediator mass $m_R$ with the DM
  mass $m_\chi = 10$~GeV. Right: Bound on the direct detection
  cross-section from LHC limits as a function of the DM mass $m_\chi$
  compared to the results from various direct detection
  experiments. The width of the blue line corresponds to the change of
  the bound as the mediator mass is varied between 300 and 1000~GeV. For larger or smaller mediator masses, the bound would become
  weaker. As an example, we show the bound obtained from the contact
  operator if the mediator can be integrated out~\cite{Fox:2011pm}.}
\label{fig:directdetection}
}
\end{center}
\end{figure}

We observe that up to $m_R \sim 1000$~GeV, $\Gamma_R / m_R$
remains sufficiently small that the NWA stays
valid, which is an important consistency requirement for our
treatment. We can therefore use the limit on $\Gamma_R$ from
Figure~\ref{fig:Gamma} to calculate an upper bound on the direct
detection cross-section. The resulting bounds on the direct detection
cross-section both as a function of mediator mass and as a function of
DM mass are shown in Figure~\ref{fig:directdetection}. We observe that
we can exclude a cross-section of $\sigma_p =
2\times10^{-40}\text{cm}^2$ over the full mass range $300\text{ GeV}
\leq m_R \lesssim 1000 \text{ GeV}$, as long as $m_\chi \ll m_R$.

\begin{figure}[t]
\begin{center}
{
\includegraphics[height=.30\columnwidth]{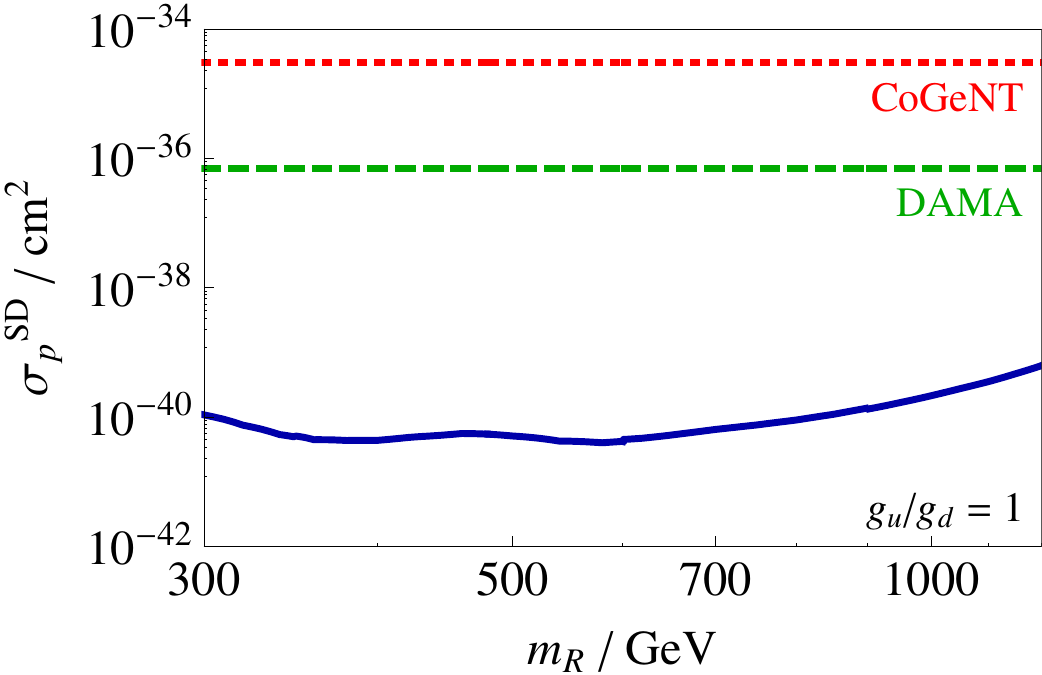}
\includegraphics[height=.30\columnwidth]{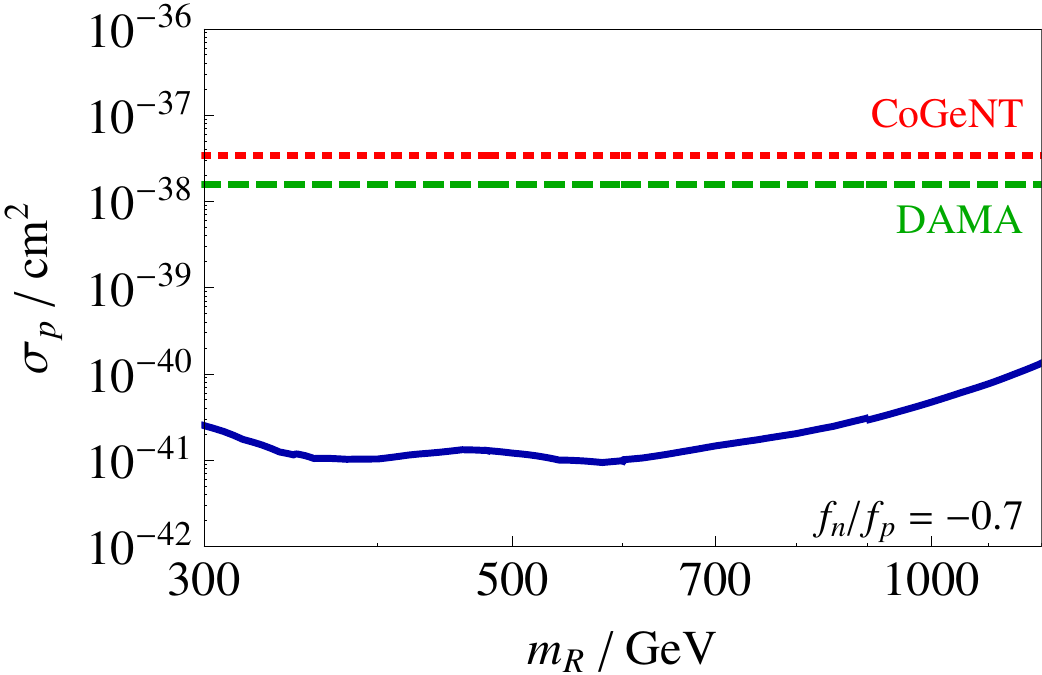}
\vspace*{-3mm}
\caption{Left: Bound on the spin-dependent direct detection
  cross-section from LHC limits as a function of the mediator mass
  $m_R$ with the DM mass fixed to $m_\chi = 10$~GeV. Right: Same but
  for spin-independent interactions with the isospin-violating
  couplings $f_n / f_p = -0.7$.}
\label{fig:directdetection2}
}
\end{center}
\end{figure}

So far, we have only considered standard spin-independent interactions of
DM. Many other possibilities have been considered in the literature,
e.g.\ spin-dependent interactions, momentum-dependent interactions,
inelastic DM, and effective couplings with $f_n \neq f_p$ (see
e.g.~\cite{Frandsen:2011ts}). Typically, these interactions strongly
suppress scattering in the non-relativistic limit, while the results
from LHC searches are not significantly affected. As a result, the LHC
bounds will become much stronger compared to any exclusion
limits (or claimed signals) from direct detection experiments. To
illustrate this point, we show in Figure~\ref{fig:directdetection2} the
LHC bounds for spin-dependent interactions, as well as
spin-independent interactions with $f_n / f_p = -0.7$.

To conclude this section, we discuss how our results would change for complex
scalar DM (e.g.~a scalar technibaryon \cite{Gudnason:2006yj,Foadi:2008qv,Barbieri:2010mn,Frandsen:2011kt}). In fact, all the experimental limits can
be applied in complete analogy, the only difference being that for identical couplings
the partial width for decays into scalar DM is smaller by a factor of
4, cf.~Appendix~\ref{sec:widths}. As a consequence, the bounds on the direct detection cross-section
will be \emph{weaker} by a factor of 4, excluding $\sigma_p > 8 \times
10^{-40} \text{ cm}^2$ for $m_\chi \ll m_R$ and $300\text{ GeV} < m_R
< 1000$~GeV.

\subsection{Direct detection through mixing}
\label{sec:ddmixing}

Now we consider the case where $R$ is the mass eigenstate that
corresponds to the gauge boson $X$ of a new $U(1)_X$ gauge group.
$X$ is then described by an effective Lagrangian, which
includes kinetic mixing and mass mixing (see also Appendix~\ref{sec:couplings})
\begin{align}
  {\cal L} =& \; {\cal L}_{SM}
  -\frac{1}{4}\hat{X}^{\mu\nu}\hat{X}_{\mu\nu} + {\frac{1}{2}} m_{\hat
    X}^2 \hat{X}_\mu \hat{X}^\mu - m_\chi \bar{\chi}\chi
  \nonumber  \\
  & - \frac{1}{2} \sin \epsilon\, \hat{B}_{\mu\nu} \hat{X}^{\mu\nu} +\delta
  m^2 \hat{Z}_\mu \hat{X}^\mu - f^\mathrm{V}_\chi \hat{X}^\mu
  \bar{\chi}\gamma_\mu \chi \;.
\label{LZprime}
\end{align}
We assume that the interaction eigenstate
$X$  couples only to the DM particle $\chi$ (assumed here to be a Dirac fermion) with strength
$f^\mathrm{V}_\chi$, and has no other couplings. In particular, we
assume that other hidden sector states~-- even if present~-- give a
negligible contribution to the total width of $R$.

The mass eigenstate
$R$ then picks up SM couplings from mixing as described in
Appendix~\ref{sec:couplings}. At the same time, kinetic mixing
introduces a coupling between $\chi$ and the $Z$-boson. After
integrating out both $R$ and $Z$, the resulting coupling constants for
the effective interactions between DM and nucleons are, in terms of
the Lagrangian parameters in Equation~(\ref{LZprime})
\begin{align}
  f_p =& \frac{\hat{g} f^\mathrm{V}_\chi}{4 \hat c_\mathrm{W} }
  \frac{c_\xi^2}{c_\epsilon} t_\xi \left[ \left(1-4{\hat
      s_\mathrm{W}}^2    -3 {\hat s_\mathrm{W}} {t_\epsilon t_\xi}\right)\frac{1}{ m_Z^2} - \left(1-4{\hat
      s_\mathrm{W}}^2    +3 {\hat s_\mathrm{W}} \frac{t_\epsilon}{ t_\xi}\right)  \frac{1}{ m_{R}^2} \right]\;,
  \nonumber\\
  f_n =& - \frac{\hat{g} f^\mathrm{V}_\chi}{4 \hat c_\mathrm{W} }
  \frac{c_\xi^2}{c_\epsilon} t_\xi 
  \left[ \left(1+ {\hat s_\mathrm{W}} {t_\epsilon t_\xi}\right)\frac{1}{m_Z^2} 
  -  \left(1-\hat s_\mathrm{W} \frac{t_\epsilon}{t_\xi}\right) 
  \frac{1}{ m_{R}^2} \right] \;,
\label{eq:fnfpmixing}
\end{align}
where $\xi$ is defined in Equation~(\ref{eq:xi}) and we abbreviated $\sin \theta \equiv s_\theta, \cos \theta \equiv c_\theta, \tan\theta \equiv t_\theta$.
In the case where the mass mixing parameter $\delta m = 0$, we obtain
$f_n \simeq 0$, i.e.\ photon-like interactions.

Because of mixing
$R$ can be produced directly in $p$-$p$
collisions, so we can use LHC data to constrain the mixing
parameters 
and therefore obtain bounds on the direct detection cross-section.
In addition to the LHC bounds, we
also have LEP bounds on the kinetic mixing parameters $\sin \epsilon$
and $\xi$ and on $g_{\chi Z}$.
In order to satisfy electroweak precision tests (EWPT) we must require
that~\cite{Babu:1997st}
\begin{align}
 \alpha S &= 4 \xi {\hat c}_\mathrm{W}^2 {\hat s}_\mathrm{W} t_{\epsilon} , \quad \text{and}\\
 \alpha T &= \xi^2 (m_{\hat X}^2/m_{\hat Z}^2-1)+2\xi {\hat s}_\mathrm{W} t_{\epsilon} ,
\end{align}
are within their experimental limits.  Moreover, from measurements of
the $Z$ invisible width, we know that 
\beq 
\left(g_{\chi Z}\right)^2
\lesssim 0.008 \; .  
\eeq 
Note, however, that new physics might well give additional
contributions to the $S$ and $T$ parameters, which can modify these bounds.

 \begin{figure}[tb]
\begin{center}
{
\includegraphics[width=.49\textwidth,clip,trim=5 0 35 0]{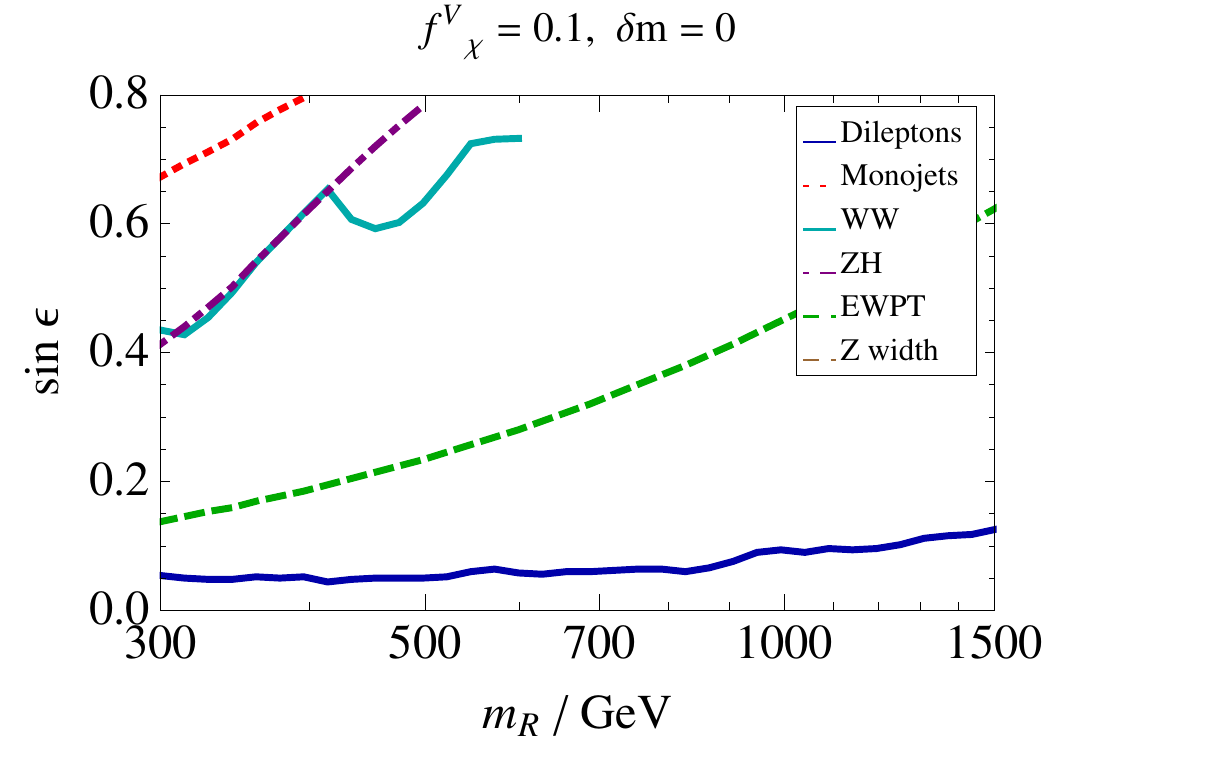}
\includegraphics[width=.49\textwidth,clip,trim=5 0 35 0]{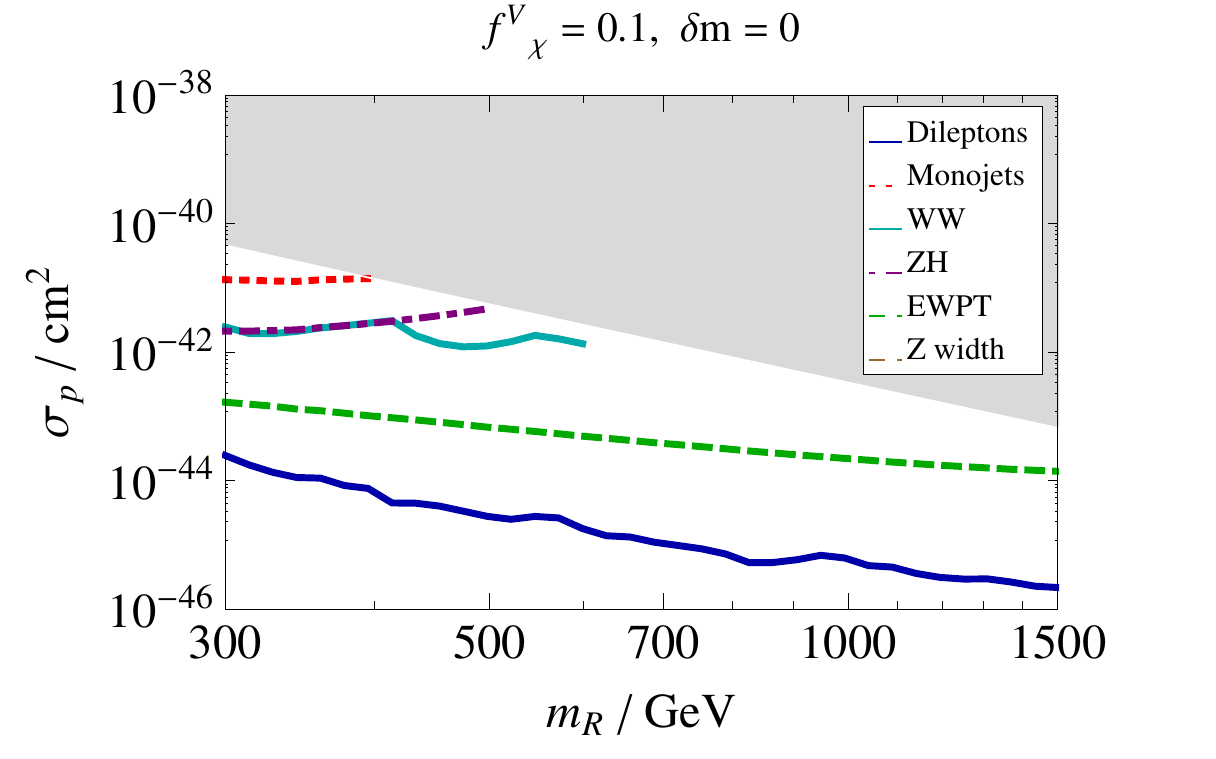}
\vspace*{-3mm}
\caption{Bounds on the mixing parameter $\sin \epsilon$ and the
  direct detection cross-section $\sigma_p$ for $f^V_\chi = 0.1$ and
  $\delta m = 0$. The grey shaded region in the second plot
  corresponds to $\sin \epsilon  > 0.8$.}
\label{fig:mixing}
}
\end{center}
\end{figure}

For fixed $f^\mathrm{V}_\chi$ and $\delta m$ 
we can calculate all
coupling constants and therefore the partial decay widths and
branching ratios of $R$ as a function of $m_R$ and
$\sin \epsilon$. The bounds from Section~\ref{sec:bounds} can then be
interpreted as constraints on $\sin \epsilon$ as a function of $m_R$. 
These constraints directly correspond to
limits on the DM direct
detection cross-section. Our results are presented in
Figure~\ref{fig:mixing}.

\begin{figure}[tb]
\begin{center}
{
\includegraphics[width=.49\columnwidth,clip,trim=5 0 35 0]{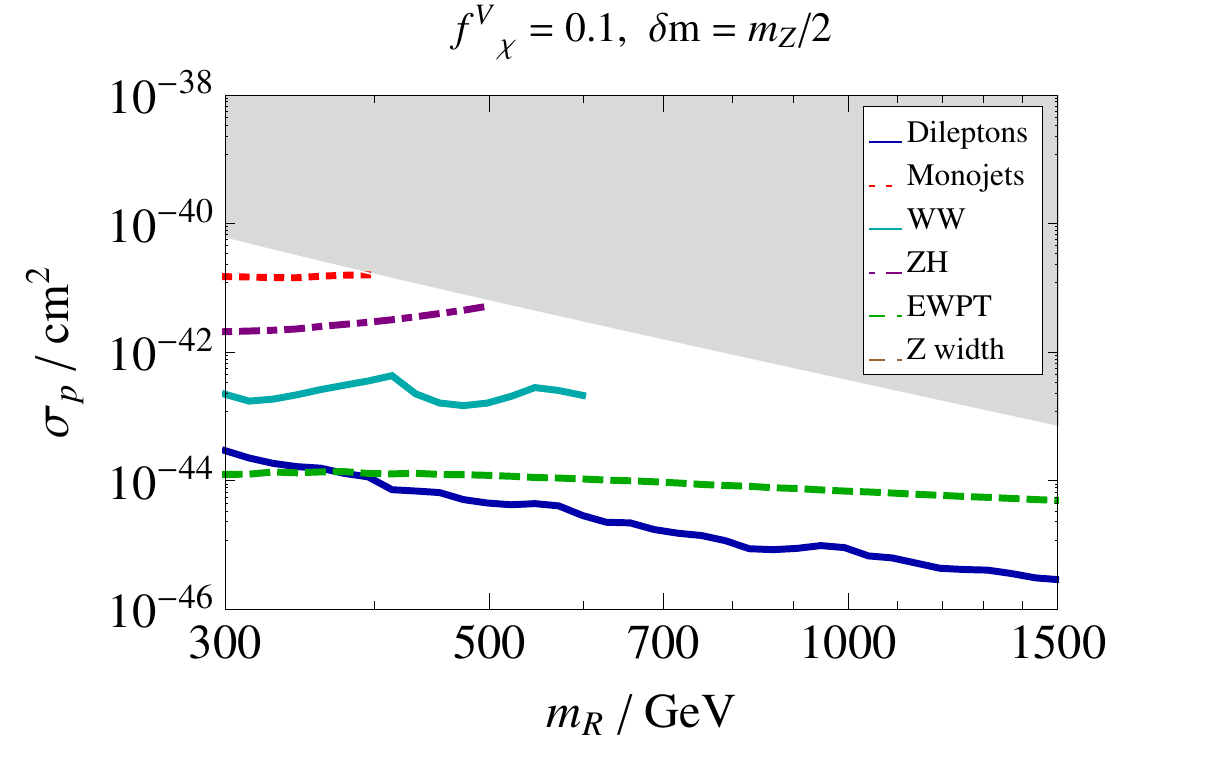}
\includegraphics[width=.49\columnwidth,clip,trim=5 0 35 0]{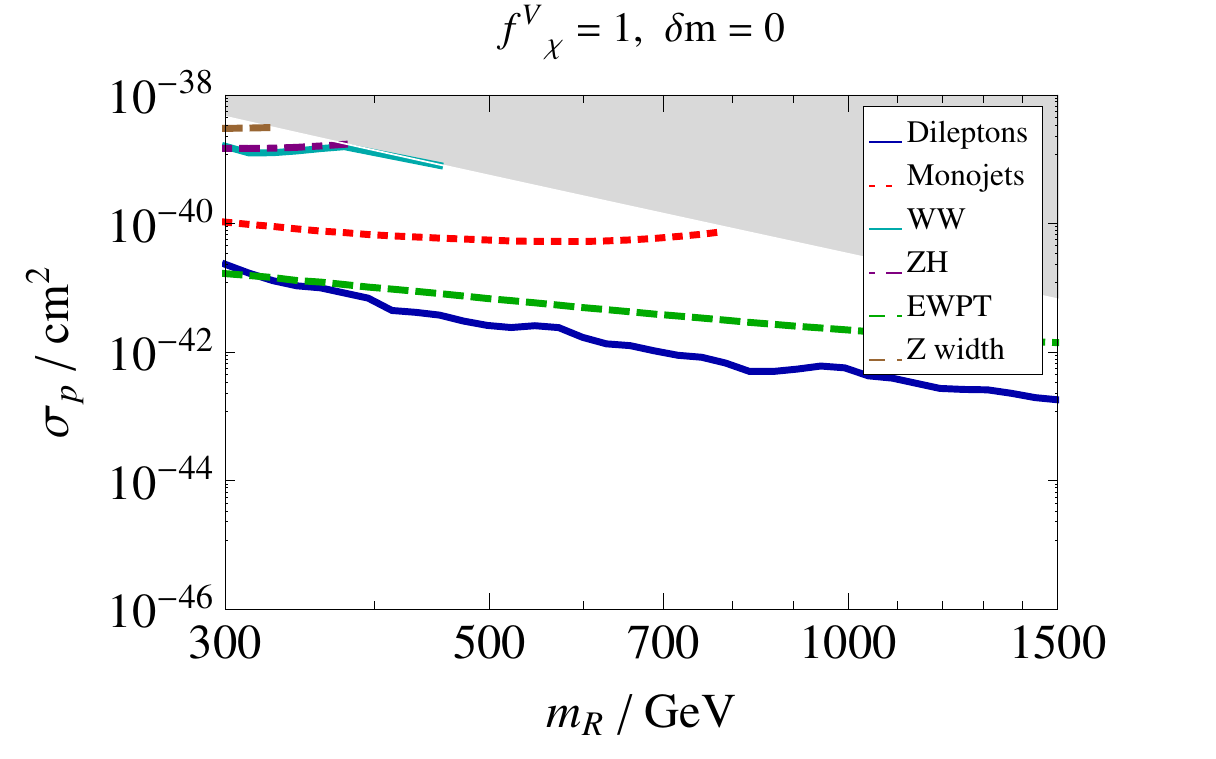}
\vspace*{-3mm}
\caption{Bounds on the direct detection cross-section for different
  choices of $f^V_\chi$ and $\delta m$. The grey shaded regions
  correspond to $\sin \epsilon  > 0.8$.}
\label{fig:mixing2}
}
\end{center}
\end{figure}

We observe that the bound from $WW$ does not get stronger compared to
the other bounds as we increase $m_R$, despite the enhancement of the
branching ratio of $R$ into $WW$. The reason is that
for small kinetic mixing, the coupling of $R$ to quarks and leptons is
proportional to $\cos \xi t_{\epsilon} \approx \epsilon$, while the
coupling of $R$ to $WW$ is proportional to $\sin \xi \approx \epsilon
\sin \theta_W M_Z^2 / m_R^2$. Thus, the partial width $\Gamma^{WW}$
picks up an additional factor of $m_Z^4 / m_R^4$ which precisely
cancels the enhancement from the derivative interaction.

In the case where $f^V_\chi = 0.1$, the LHC gives strong constraints
on the direct detection cross-section, comparable to the best bounds
from current direct detection experiments. This conclusion does not
change significantly, if we include a mass mixing term (see
Figure~\ref{fig:mixing2}). However, such a mass mixing will enhance
$\xi$ compared to $\epsilon$, so the bound from $WW$ becomes stronger
compared to the bound from dileptons and monojets.

Increasing $f^V_\chi$ relaxes all bounds from the LHC
since smaller quark couplings, and therefore smaller mixing parameters
are required for the same direct detection cross-section. At the same
time the invisible partial width of $R$ is increased so that decays of
$R$ into SM particles are additionally suppressed. Still, even for
$f^V_\chi = 1$, we can exclude a direct detection cross-section above
$10^{-41}\text{ cm}^2$, because larger mixing parameters are excluded
by both dilepton and monojet constraints as well as EWPT.

\section{Discussion}
\label{sec:discussion}

In this paper, we have considered a new vector
state $R$, as the dominant mediator of the scattering between DM and nuclei, relevant for direct detection experiments.
We have
demonstrated that the LHC can significantly constrain the scattering
cross-sections over a wide range of mediator masses. In the case of
fermionic DM and assuming that the vector mediator has direct
couplings to quarks, we can exclude cross-sections $\sigma_p > 2\times
10^{-40} \text{ cm}^2$ for $300\text{ GeV} \leq m_R \leq 1000$~GeV as
long as $m_\chi \ll m_R$, with similar limits for scalar DM. When couplings to quarks are
introduced only via mixing, the resulting bounds on the direct
detection cross-section can be much stronger.  
However, it is possible to
suppress the collider constraints by making the DM coupling very large compared to the quark couplings.
Our approach allows a very general interpretation of the results from MET searches at the LHC,
because 
we have not assumed that the mediator is heavy enough 
for an effective operator analysis. Nevertheless, we have made other
assumptions in our analysis, which we discuss below.

First, the DM particle has been assumed to have a mass $m_\chi \ll m_R$. 
When $m_\chi$ becomes larger than $m_R/2$,
$R$ can no longer decay into DM particles and therefore monojet
signals can only arise when the mediator is produced off-shell. 
Hence bounds from monojets become much weaker. 
We do not consider this
case further, as LHC bounds on the interactions of heavy DM are
significantly less stringent than those obtained from direct detection
experiments.

Second, we have assumed family
independent fermion couplings (but we allow differences between up-quark
and down-quark couplings, and between charged lepton and neutrino
couplings). Coupling the mediator predominantly to the third
generation would slightly relax the bound from dileptons,
but make the already quite strong bound on
top-quarks even stronger. At the same time, doing so would
strongly suppress the direct detection cross-section. Consequently,
allowing family-dependent couplings would not significantly change our
conclusions.

Third, we have only considered the contribution from DY production of $R$. This is
certainly justified in the case where couplings of $R$ to standard
model particles arise from kinetic mixing. For large $m_R$, however,
the coupling of $R$ to $WW$ is not well constrained, so it could
in general be possible to enhance the cross-section of VBF
production. By neglecting additional contributions to the
production of $R$ we give of course a more conservative bound.

Fourth, there is only a relatively limited range of mediator masses,
viz.~300~--~600~GeV, where we have bounds on all possible SM decay
channels. Below 300~GeV, it will be difficult to constrain some of the
couplings of $R$ using LHC data, mostly because of QCD
backgrounds. Instead, EWPT will become more important
(see~\cite{Frandsen:2011cg}). To extend the range of mediator masses
above 600~GeV, we have simply assumed that the bound on $\text{BR}(R \rightarrow WW)$ above 600~GeV is identical to that on 
$\text{BR}(R \rightarrow ZZ)$.
Since the total width of the mediator is not very
sensitive to this bound, we expect our result to be only mildly
affected if the observed bound is significantly weaker.

Finally, and most importantly, we have assumed  that we can
treat $R$ as a narrow resonance and we make extensive
use of the NWA to separate the production of
$R$ from the subsequent decays. This factorisation 
is not valid for broad resonances, where we expect a
significant contribution from off-shell mediators
(see~\cite{Accomando:2010fz} for a discussion). 
Moreover, most bounds only apply for a narrow resonance.
For example, limits for dijet resonances have only been
published for peaks with $\Gamma / M \lesssim 0.3$. 

We observe that for $m_R < 1000$~GeV, present experimental results give the
constraint $\Gamma_R/m_R < 0.25$. For these values, we expect
the NWA to be accurate within a few percent. As $m_R$ increases, the
bounds on the individual coupling constants, and therefore the bound on
$\Gamma_R$, become weaker. For $m_R > 1$ TeV, couplings can be
of order unity for which the width becomes so large that neither
experimental limits nor the formalism presented in this paper can be
applied. With increasing luminosity at the LHC, we expect to be able
to extend our treatment to larger mediator masses, while at the
same time obtaining stronger bounds on the direct detection
cross-section.

For dark matter masses $m_\chi \ll m_R$, dark matter annihilation will only involve off-shell mediators. If the mediator couples mostly to quarks and dark matter, it is therefore possible to describe annihilation in terms of the same effective operators as direct detection. Consequently, one can directly translate between direct detection cross-sections and annihilation cross sections (see for example~\cite{Zheng:2010js, Cheung:2011nt}). In the case of a vector mediator and light dark matter, direct detection cross-sections of $10^{-40}\text{ cm}^2$ or smaller correspond to an annihilation cross section significantly below the thermal cross section~\cite{Beltran:2008xg,Bai:2010hh}.

The LHC bounds on the decay channels of $R$ therefore imply that the annihilation cross section of dark matter into quarks is well below the present experimental sensitivity for dark matter indirect detection in all relevant channels including the antiproton flux from PAMELA~\cite{Adriani:2010rc}, diffuse gamma rays from FERMI-LAT~\cite{Ackermann:2011wa, GeringerSameth:2011iw, Mazziotta:2012ux} and neutrinos from the sun~\cite{Kappl:2011kz}.\footnote{However, if $R$ has only small coupling to quarks, there could be a monoenergetic gamma ray signal from annihilation of dark matter into Z gamma as in~\cite{Dudas:2012pb}.} Even more restrictive experimental bounds in the future can be evaded either if dark matter is asymmetric or by appealing to additional uncertainties as in~\cite{Charbonnier:2011ft, Cholis:2012am}.

To obtain the required dark matter relic density in our framework, we must assume either the presence of additional mediators in the early universe in order to avoid overproduction of DM or additional couplings of $R$ to new hidden sector states. In fact, coupling $R$ to new hidden sector states that decay into SM particles with more complicated experimental signatures is an interesting possibility to evade experimental limits on $\Gamma_R$. Thus there are good reasons to carry out experimental searches (see~\cite{Aad:2012vn} for an example) for such states.

\section*{Acknowledgements}

We thank Alan Barr, Georgios Choudalakis, James Unwin, Stephen West
and Stephen Worm for helpful discussions. We have also benefitted from
discussions with participants at the workshop on {\it ``New Paths to
  Particle Dark Matter''} at Oxford, 28-29 April 2012. MTF is
supported by an STFC grant, FK by the DAAD, and KSH by ERC Advanced
Grant (BSMOXFORD 228169). We also acknowledge support from the UNILHC
network (PITN-GA-2009-237920) and an IPPP associateship for 2011-12
awarded to SS.

\begin{appendix}

\section{Decay widths}
\label{sec:widths}
\begin{eqnarray}
\label{eq:widthchi}
\Gamma(R\rightarrow \chi\bar{\chi})&=&\frac{m_R}{12\pi} 
 \sqrt{1-\frac{4 m_\chi^2}{m_R^2}}  [(g^{V}_{\chi})^2+(g^{A}_{\chi})^2 + \frac{m_\chi^2}{m_R^2}(2 (g^{V}_{\chi})^2 - 4 (g^{A}_{\chi})^2)  ]  
\end{eqnarray}

\begin{eqnarray}
\Gamma(R\rightarrow \phi \phi^*)&=&\frac{m_R}{48\pi} 
 \sqrt{1-\frac{4 m_\phi^2}{m_R^2}}  g^{2}_{R\phi}  
\end{eqnarray}

\begin{eqnarray}
\Gamma(R\rightarrow f\bar{f})&=&\frac{m_R N_c}{12\pi} 
 \sqrt{1-\frac{4 m_f^2}{m_R^2}}  [(g^{V}_{f})^2+(g^{A}_{f})^2 + \frac{m_f^2}{m_R^2}(2 (g^{V}_{f})^2 - 4 (g^{A}_{f})^2)  ]  
\end{eqnarray}

\begin{eqnarray}
\Gamma(R\rightarrow W^{+}W^{-})&=&
\frac{1}{192  \pi}m_R \left (\frac{m_R}{m_W} \right )^4
\left (1-4\frac{m_W^2}{m_R^2} \right )^{1/2} \nonumber \\
&\times &\bigg((g_{WW1}^R)^2\left [4\frac{m_W^2}{m_R^2}  -4\frac{m_W^4}{m_R^4} - 48\frac{m_W^6}{m_R^6} \right ]\nonumber \\
&&+ (g_{WW2}^R)^2 \left [ 1- 16 \frac{m_W^4}{m_R^4} \right ]\nonumber \\
&&+ g_{WW1}^R g_{WW2}^R \left [12 \frac{m_W^2}{m_R^2} - 48\frac{m_W^4}{m_R^4}  \right ]\nonumber \\
&&+ (g_{WW3}^R)^2 \left [4 \frac{m_W^2}{m_R^2}- 32 \frac{m_W^4}{m_R^4} + 64 \frac{m_W^6}{m_R^6}  \right ] \bigg)
\end{eqnarray}

\begin{eqnarray}
\Gamma(R\rightarrow Z Z )&=&
\frac{(g_{ZZ}^R)^2}{96  \pi}m_R \frac{m_R^2}{m_Z^2}
\left (1-4\frac{m_Z^2}{m_R^2} \right )^{3/2}
\left [1 - 6\frac{m_Z^2}{m_R^2} \right ]
\end{eqnarray}

\begin{eqnarray}
\Gamma(R\rightarrow Z \gamma )&=&
\frac{(g_{Z\gamma}^R)^2}{96  \pi}m_R \frac{m_R^2}{m_Z^2}
\left (1-\frac{m_Z^2}{m_R^2} \right )^{3}
\end{eqnarray}

\begin{eqnarray}
\Gamma(R\rightarrow ZH)&=&\frac{(g^R_{ZH})^2}{192\pi m_Z^2} m_R 
\sqrt{\lambda(1,x_Z, x_H)}(\lambda(1,x_Z, x_H)+ 12 x_Z) \ , 
\end{eqnarray}
where $x_Z=(m_Z/m_R)^2$, $x_H=(m_H/m_R)^2$, and
$\lambda(x,y,z)=x^2+y^2+z^2-2xy-2yz-2zx$. Note that in the latter
formula $g^R_{ZH}$ has mass dimension one, following our conventions
in the main text, and the expression agrees with that given in
e.g.~\cite{Barger:2009xg}.

\section{Coupling structure from mixing}
\label{sec:couplings}

In this appendix, we discuss how the mass eigenstate $R$ arises from the mixing of an
interaction eigenstate vector $X$ with the SM $U(1)_Y$ $B$ field and
the neutral component $W^3$ of $SU(2)_\mathrm{L}$ weak
fields. We first consider the most general case and calculate the effective coupling constants defined in Section~\ref{sec:theory} in terms of the fundamental couplings and the entries of the mixing matrix. We then calculate the mixing matrix that arises from kinetic mixing and mass mixing of gauge bosons. 

Following the notation in~\cite{Frandsen:2011cg} we write
the general mixing matrix as
\begin{align}
\label{eq:general-mixing}
\left(\begin{array}{c} \hat{B}_\mu \\ \hat{W}^3_\mu \\ \hat{X}_\mu \end{array}\right)=
\left(\begin{array}{ccc}  N_{11} & N_{12} & N_{13}
\\ N_{21} & N_{22} & N_{23}
\\ N_{31} & N_{32} & N_{33}
\end{array}\right)
\left(\begin{array}{c} A_\mu \\ Z_\mu \\ R_\mu \end{array}\right).
\end{align}
Here $A$, $Z$ are the physical photon and neutral massive gauge boson
fields of the SM.
The couplings of $R$ to SM fermions are given in terms of the mass
mixing matrix, as\footnote{We use $\hat{g}$ and
$\hat{g}'$ to denote the fundamental gauge couplings of 
$SU(2)_\text{L}$ and $U(1)_Y$, which will be different from the observed ones, $g$ and
$g'$.}
\begin{align}
  g_{u R}^\mathrm{V} &= -\frac{1}{12}(5 {\hat g'} N_{13}+3 {\hat
    g}N_{23})-f_u^\mathrm{V} N_{33} \ , & g_{u R}^\mathrm{A} &=
  \frac{1}{4}({\hat g'} N_{13} - {\hat g}N_{23})-f_u^\mathrm{A} N_{33}
  \ ,
  \nonumber \\
  g_{d R}^\mathrm{V} &= \frac{1}{12}({\hat g'} N_{13}+3 {\hat
    g}N_{23})-f_d^\mathrm{V} N_{33} \ , & g_{d R}^\mathrm{A} &= -\frac{1}{4}
  ({\hat g'} N_{13} - {\hat g}N_{23})-f_d^\mathrm{A} N_{33}\ ,
  \nonumber \\
  g_{e R}^\mathrm{V} &= \frac{1}{4}(3 {\hat g'} N_{13}+ {\hat
    g}N_{23})-f_e^\mathrm{V} N_{33} \ , & g_{e R}^\mathrm{A} &= -\frac{1}{4}
  ({\hat g'} N_{13} - {\hat g}N_{23})-f_e^\mathrm{A} N_{33}\; ,
    \nonumber \\
  g_{\nu R}^\mathrm{V} &= \frac{1}{4}( {\hat g'} N_{13}- {\hat
    g}N_{23})-f_\nu^\mathrm{V} N_{33} \ , & g_{\nu R}^\mathrm{A} &= -\frac{1}{4}
  ({\hat g'} N_{13} - {\hat g}N_{23})-f_\nu^\mathrm{A} N_{33}\; ,
\label{couplings}
\end{align}
where the numerical coefficients are determined from the hypercharge
and weak quantum numbers of the SM fermions and $f^\mathrm{V, A}$
denote the direct couplings of $X$. 
Similarly, the effective vector and axial couplings of $R$ to
the DM particle are given by
\begin{align}
  g_{\chi R}^\mathrm{V} = f_\chi^\mathrm{V} N_{33} \ , & \ \ 
  g_{\chi R}^\mathrm{A} = f_\chi^\mathrm{A} N_{33} \  & \text{or} & &
  g_{\phi R} = & f_\phi N_{33} \ \;
\label{DMcouplings}
\end{align}
depending on whether the DM particle is a fermion or a scalar.

Finally, the couplings of $R$ to SM bosons are given by
\cite{Chun:2010ve}
\begin{align}
&g^{R}_{WW1} = g^{R}_{WW2} = \hat{g} N_{23} \; , \quad g^{R}_{ZWW1} =
-\hat{g}^2 N_{22} N_{23} \; , \quad g^{R}_{AWW1} = - \hat{g}^2 N_{21}
N_{23} \; , \\ 
&g^{R}_{ZH} = \frac{v}{2}({\hat g'} N_{12} - {\hat g}
N_{22}) ({\hat g'} N_{13} - {\hat g} N_{23}) \; , \ g^{R}_{ZHH} =
\frac{1}{2}({\hat g'} N_{12} - {\hat g} N_{22}) ({\hat g'} N_{13} -
     {\hat g} N_{23}).
\label{WWcouplings}
\end{align}
Since $P$-violating couplings of gauge bosons are absent in the SM, the
corresponding couplings of $R$ cannot be introduced by mixing alone.

We now assume that
$X$ is the gauge boson of a new $U(1)_X$ gauge group and follow the
discussion in~\cite{Frandsen:2011cg} of an effective Lagrangian which
includes kinetic mixing and mass mixing (see also~\cite{Babu:1997st})
\begin{align}
  {\cal L} =& \; {\cal L}_{SM}
  -\frac{1}{4}\hat{X}^{\mu\nu}\hat{X}_{\mu\nu} + {\frac{1}{2}} m_{\hat
    X}^2 \hat{X}_\mu \hat{X}^\mu - m_\chi \bar{\chi}\chi
  \nonumber  \\
  & - {\frac{1}{2}} \sin \epsilon\, \hat{B}_{\mu\nu} \hat{X}^{\mu\nu} +\delta
  m^2 \hat{Z}_\mu \hat{X}^\mu - \sum_{f} f_f^\mathrm{V} \hat{X}^\mu
  \bar{f}\gamma_\mu f - f^\mathrm{V}_\chi \hat{X}^\mu
  \bar{\chi}\gamma_\mu \chi \;.
\label{eq:Lappendix}
\end{align}
As in~\cite{Weihs:2011wp}, we assume that the $U(1)_X$ is broken by an
additional Higgs field and $X$ acquires the mass $m_{\hat X}$. We will
not discuss the implications of this additional Higgs field and its
potential mixing further. 
We define $\hat{Z}\equiv 
\hat{c}_\mathrm{W} \hat{W}^3- \hat{s}_\mathrm{W} \hat{B}$, where $\hat{s}_\mathrm{W} \, (\hat{c}_\mathrm{W})$ is the sine (cosine) of the (fundamental) Weinberg angle. 

The diagonalisation of the above Lagrangian is discussed in detail in
e.g.~\cite{Babu:1997st}.  The field strengths are diagonalised and
canonically normalised via the following two consecutive
transformations
\begin{align}
\label{eq:Zpmixing}
\left(\begin{array}{c} \hat B_\mu \\ \hat W_\mu^3 \\ \hat
    X_\mu \end{array}\right) & = \left(\begin{array}{ccc} 1 & 0 &
    -t_\epsilon \\ 0 & 1 & 0 \\ 0 & 0 &
    1/c_\epsilon \end{array}\right)
\left(\begin{array}{c} B_\mu \\ W_\mu^3 \\ X_\mu \end{array}\right) \ , \\
\left(\begin{array}{c} B_\mu \\ W_\mu^3 \\ X_\mu \end{array}\right) &
= \left(\begin{array}{ccc}
    \hat c_\mathrm{W} & -\hat s_\mathrm{W} c_\xi &  \hat s_\mathrm{W} s_\xi \\
    \hat s_\mathrm{W} & \hat c_\mathrm{W} c_\xi & - \hat c_\mathrm{W} s_\xi \\
    0 & s_\xi & c_\xi
\end{array} \right) 
\left(\begin{array}{c} A_\mu \\ Z_\mu \\ R_\mu \end{array}\right)
 \; ,
\end{align}
where
\begin{align}
  t_{2\xi}=\frac{-2c_\epsilon(\delta m^2+m_{\hat Z}^2 \hat
    s_\mathrm{W} s_\epsilon)} {m_{\hat X}^2-m_{\hat
      Z}^2 c_\epsilon^2 +m_{\hat Z}^2\hat s_\mathrm{W}^2 s_\epsilon^2
    +2\,\delta m^2\,\hat s_\mathrm{W} s_\epsilon} \; .
\label{eq:xi}
\end{align}
Multiplying the two matrices, we obtain the coefficients $N_{ij}$ so
that we can calculate the couplings of $R$ using
Equations~(\ref{couplings}--\ref{WWcouplings}). As discussed
in~\cite{Chun:2010ve,Frandsen:2011cg}, the fundamental parameters
$m_{\hat Z}$ and $\hat{s}_\mathrm{W}$ are constrained by the requirement that
the physical $Z$ mass and the Weinberg angle come out in accord with experiment.

\begin{figure}[!ht]
\begin{center}
{
\includegraphics[width=.49\columnwidth]{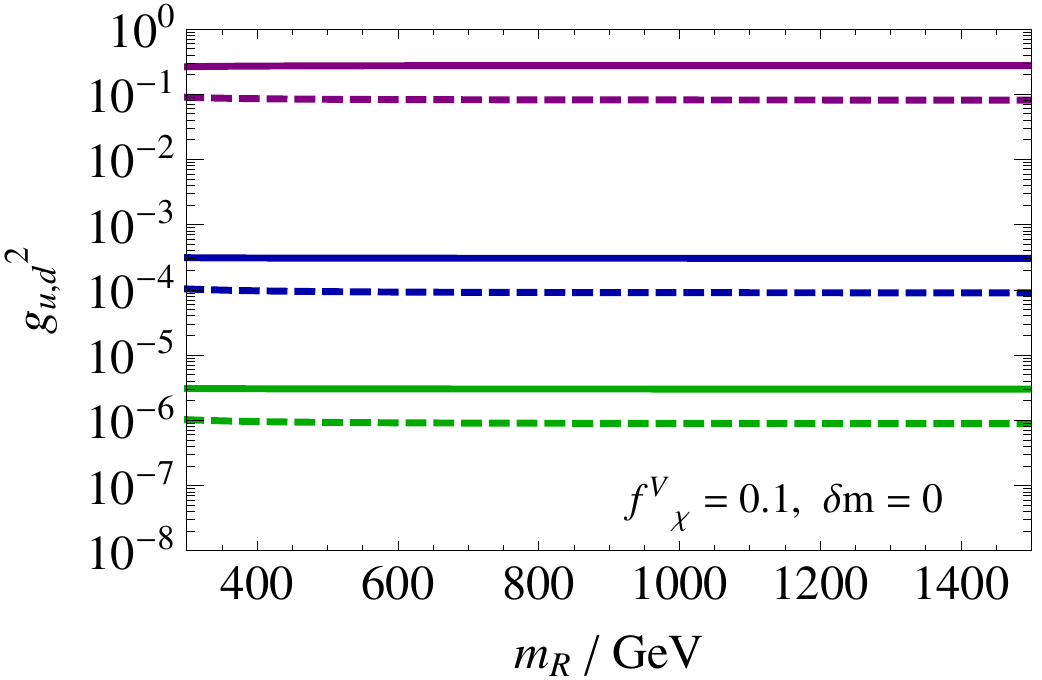}
\includegraphics[width=.49\columnwidth]{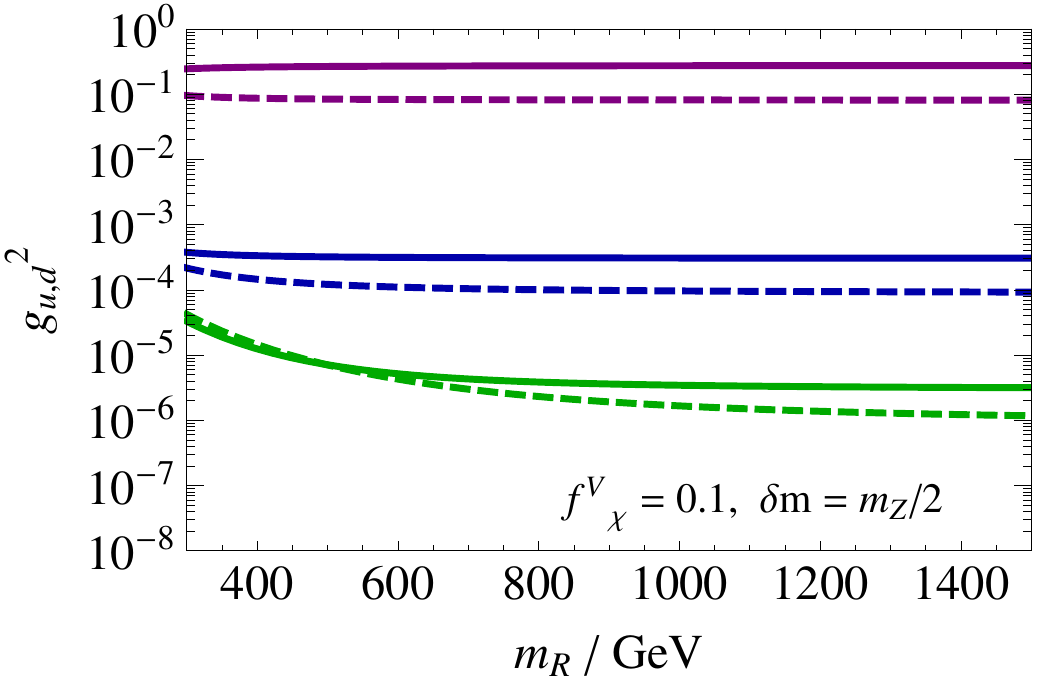}
\vspace*{-3mm}
\caption{The couplings $g_{u}^2$ (solid lines) and $g_{d}^2$ (dashed
  lines) as a function of $m_R$ for $\epsilon=0.01$ (green), $\epsilon
  = 0.1$ (blue) and $\epsilon=1$ (purple). In the left plot,
  interactions are induced by kinetic mixing only via the Lagrangian
  given in Equation~(\ref{eq:Lappendix}), while in the right plot we have included
  a mass mixing of $\delta m = m_Z/2$.  }
\label{fig:gud2plots}
}
\end{center}
\begin{center}
{
\includegraphics[width=.49\columnwidth]{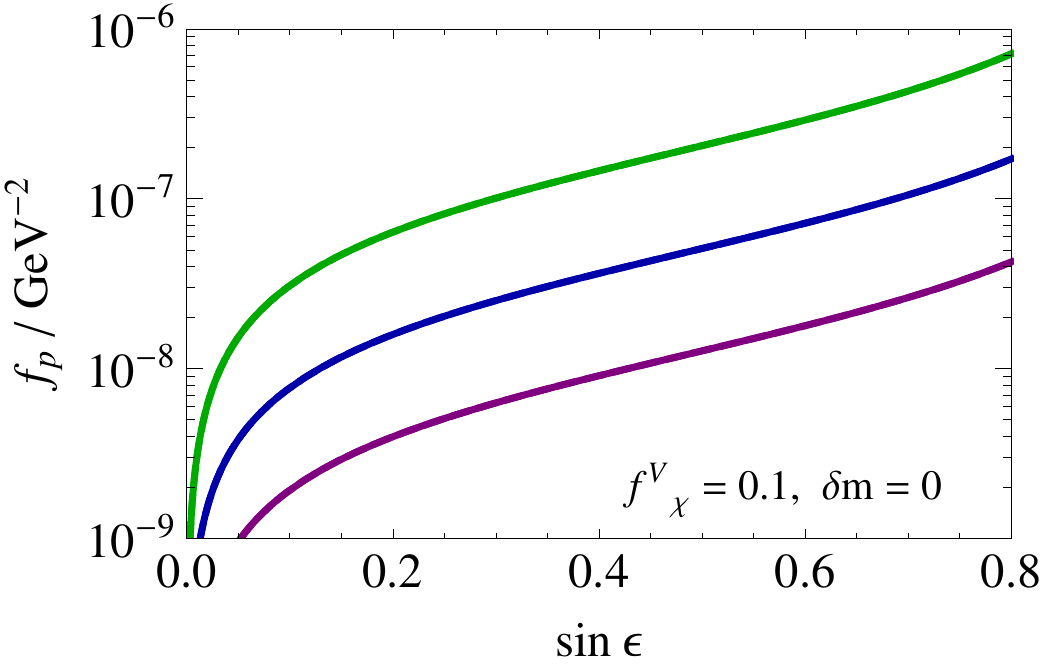}
\includegraphics[width=.49\columnwidth]{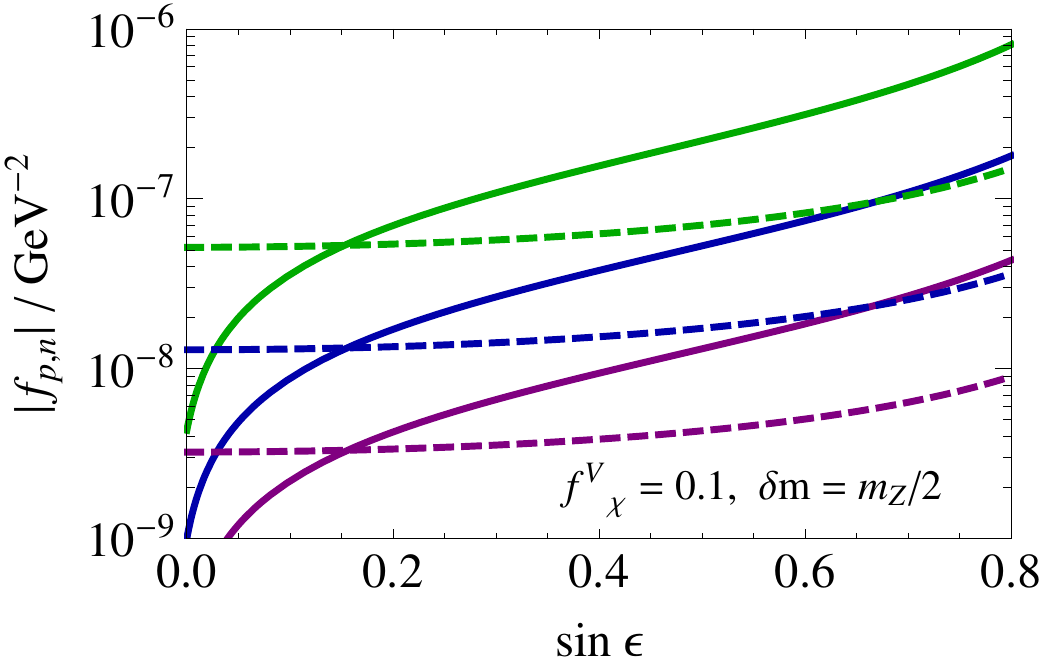}
\vspace*{-3mm}
\caption{The effective couplings $f_p$ (solid lines) and $f_n$ (dashed
  lines) obtained from the Lagrangian given in Equation~(\ref{eq:Lappendix}) 
  as a function of $\sin \epsilon$ for $m_R = 300$~GeV (green), $m_R
  = 600$~GeV (blue) and $m_R = 1200$~GeV (purple). In the left plot,
  interactions are induced by kinetic mixing only, so $f_n = 0$, while in the right plot we
  have included a mass mixing of $\delta m = m_Z/2$.  }
\label{fig:fpnplots}
}
\end{center}
\end{figure}

At colliders we are
directly sensitive to the couplings $g_{u,d}$, which determine the
production cross-section of $R$ (see Section~\ref{sec:production}). We
show these in Figure~\ref{fig:gud2plots} as a function of $m_R$ for
different values of the kinetic mixing parameter $\epsilon$ and
different values of the mass mixing parameter $\delta m$.

If there are no direct couplings to quarks, we can calculate the effective DM-nucleon couplings $f_p$ and $f_n$ in terms of $\epsilon$, $\delta m$ and $f^V_\chi$, cf.~Equation~(\ref{eq:fnfpmixing}).
 We show these couplings as a function of $\sin \epsilon$ for different values of $\delta m$ in Figure~\ref{fig:fpnplots}. Note that for $\delta m = 0$, we obtain $f_n/f_p = 0$, while for $\delta m \neq 0$, $f_n / f_p < 0$.

\end{appendix}

\providecommand{\bysame}{\leavevmode\hbox to3em{\hrulefill}\thinspace}

\end{document}